\documentclass[
superscriptaddress,
bibnotes,
 amsmath,amssymb,
 aps,
prd,
floatfix,
]{revtex4-2}
\usepackage{amsmath,amssymb,amsfonts,mathrsfs,bm,bbold,braket,slashed,cancel}
\usepackage{graphicx}
\usepackage{epsfig}
\usepackage{float}
\usepackage{subfigure}
\usepackage{subcaption}
\usepackage{caption}
\usepackage[labelformat=simple]{subcaption}
\usepackage{multirow}
\usepackage{comment}
\usepackage[usenames,dvipsnames,svgnames]{xcolor}
\usepackage{hyperref}
\usepackage{etoolbox}
\usepackage{dsfont}
\usepackage{orcidlink}
\hypersetup{
    colorlinks=true,
    linkcolor=Blue,
    citecolor=Blue,
    urlcolor=Blue
}
\usepackage{accents}
\usepackage[normalem]{ulem}
\usepackage{simpler-wick}

\newcommand{\bea}{\begin{eqnarray}}
\newcommand{\eea}{\end{eqnarray}}
\newcommand{\ee}{\end{equation}}
\newcommand{\be}{\begin{equation}}

\begin{document}

\title{Two gluon fusion into a $\pi^0$, $\eta$, or $\eta'$ in a relatively weak magnetic field}

\author{Fabio L. Braghin\orcidlink{0000-0002-2321-2167}
}

\email{ braghin@ufg.br }

\affiliation{Instituto de F\'isica, Universidade Federal de Goi\'as,  Av. Esperança, s/n, 74690-900,  Goiânia, GO,  Brazil}

\author{Enrique Mu\~noz~\orcidlink{0000-0003-4457-0817}}
 \email{ejmunozt@uc.cl}
 \affiliation{Facultad de Física, Pontificia Universidad Católica de Chile, Vicuña Mackenna 4860, Santiago, Chile}

\begin{abstract}
In this article, we investigate gluon fusion as a mechanism for neutral meson production ($\pi^0$, $\eta$, and $\eta'$), in the phenomenological context of heavy-ion collisions prior to thermalization, in which a large gluon multiplicity is expected, and when an intense background magnetic field is expected to arise. For this purpose, we computed the ratios of the emission probability rates in the presence of the background magnetic field as compared to vacuum (zero-field), as a function of transverse momenta $p_T$ and energy $\sqrt{s}$, close to midrapidity. Our theoretical results suggest a catalytic effect from the presence of the background magnetic field, that may enhance the neutral pion production by several orders of magnitude at very small energies $s \lesssim  1~\text{GeV}^2$, while the effect is negligible at larger energies $s\gtrsim 10~\text{GeV}^2$. 
In contrast, the $\eta$ and $\eta'$ production rates are nearly insensitive to the magnetic field presence
 (as compared to vacuum).
\end{abstract}

\maketitle

\section{Introduction}

Heavy ion collisions (HIC) at relativistic momenta~\cite{Andronic_2018,Busza_2018} constitute our state of the art experimental approach to test the properties of the quark gluon plasma (QGP)~\cite{McLerran_RevModPhys.58.1021,SATZ_2011,Blaizot_2002}. This state of matter is predicted by Quantum Chromodynamics (QCD) to emerge above a critical energy density, that was presumably attained a few microseconds after the Big-Bang. In HIC, colliding nuclei generate small systems of very high energy density, allowing to explore the phase diagram of QCD, particularly the smooth crossover transition leading to a deconfined phase with chiral restoration.~\cite{McLerran_RevModPhys.58.1021,Blaizot_2002,BANKS1980103,Andronic_2018,Busza_2018,Acharya_2018}

Among the many possible observables to probe the QGP, the detection of $\pi^0$ and $\eta$ meson production over a wide range of transverse momenta ($p_T$) in different colliding nuclei is of particular interest. These two neutral mesons are the most abundant sources for photon production by decay processes, and hence they constitute the primary background for the study of direct photons and heavy-flavor measurements. Interestingly, the measurements of charged pion spectra at low $p_T$ from RHIC~\cite{Adare_PHENIX_2013,Adler_STAR_2001} are compatible with thermal models~\cite{Broniowski_PRL_2001} 
assuming a hydrodynamic expansion near local equilibrium, thus leading to the interpretation of a thermalized system of quarks and gluons formed at early stages of the collision. Moreover, LHC data at low $p_T$ are compatible~\cite{Acharya_2018} with a sudden hadronization of the QGP, leading to the emergence of additional non-equilibrium chemical potentials for light and strange quarks, as captured by equilibrium and chemical non-equilibrium statistical hadronization models~\cite{Begun_PRC_2014,Begun_PRC_2015,Abuki_PRD_2009,MIGDAL_PLB_1979} (EQ SHM and NEQ SHM). These effects can be interpreted as the onset of "pion condensation" in ultra-relativistic HIC at these energies. The ALICE collaboration~\cite{Acharya_2018} at the Large Hadron Collider (LHC) has measured pion and $\eta$ meson production, in the transverse momentum range $1 < p_T < 20$~GeV/c and at midrapidity, in central and semi-central $Pb+Pb$ collisions, at energies (per nucleon) $\sqrt{s_{NN}} = 2.76$~TeV. The same group also measured $\eta$ meson production, reporting a ratio $\eta/\pi^0 = 0.457 \pm 0.013^{\rm{stat}} \pm 0.018^{\rm{syst}}$ for high $p_T$~\cite{Acharya_2018}. Interestingly, in the range $6 < p_T < 10$~GeV a stronger relative suppression is reported~\cite{Acharya_2018}, with experimental data exhibiting some tension with the hydrodynamic and pQCD hadronization models for $p_T < 3$~GeV, but no experimental data is available yet in the very small range $p_T < 1$~GeV. It has been proposed that a possible, alternative mechanism for meson production, in proton-proton collisions, is via pomeron-pomeron fusion processes~\cite{lebiedowicz_2026} in the region $\sqrt{s} = 29.1$~GeV, and $\sqrt{s} = 14$~TeV, respectively, but this models also exhibit some discrepancy with experimental data~\cite{Barberis_1998} in the low $p_T$ and low scattering angle regions.

It is widely believed that, given the charge distribution in the nucleons participating in non-central HIC, a very intense local magnetic field~\cite{Inghirami_2020,KHARZEEV_NPA_2008,Skokov_IJMPA_2009,ADHIKARI_2026,Toneev_2016} (on the order of $|e B| \sim m_{\pi}^2 \sim 10^{14}$~T), should emerge within the collision region, lasting for a few femtoseconds~\cite{Huang_dynamicalMF_2023}. 
Moreover, measurements of charge-dependent directed flow for charged mesons at RHIC~\cite{Abdulhamid_PRX_2024}, in $Au + Au$ collisions at $\sqrt{s_{NN}} = 200\,$~GeV and at $\sqrt{s_{NN}} = 27\,$~GeV, as measured by the STAR detector~\cite{Abdulhamid_PRX_2024}, provides direct experimental evidence for the existence of those intense electromagnetic fields. Therefore, one may expect that the presence of this intense magnetic field should influence the different mechanisms of hadron production, that necessarily involve the participation of charged quark species, even for neutral mesons. In consequence, a large body of theoretical work involving the interaction of charged fermions with this external magnetic field, whose foundations start with Schwinger's seminal work~\cite{schwinger1951gauge} on the proper-time formalism, has been reported in the literature~\cite{ADHIKARI_2026,Fukushima_PRD_2011,Jancovici_PR_1969,Machet_IJMPA_2016,Ayala_Munoz_2021}, even including the effects of statistical fluctuations in the background field~\cite{Munoz_Castano_2025,Munoz_Castano_2024,Munoz_Castano_PRD_2024,Munoz_Castano_2023}.  Despite gluons and neutral mesons are insensitive to magnetic fields at tree-level, higher-order corrections involving vacuum fluctuations with virtual charged quark species may also induce interactions with those fields~\cite{Heisenberg_1,DUNE,Ayala_PRD_2017,Ayala_PRD_2020}.
 Although these magnetic fields are extremely intense, they may be relatively weak when compared to an hadronic energy scale, such as 
a constituent quark mass ($M$),
i.e. $|eB| < M^2$.

Among the different mechanisms leading to "pion condensation" in the QGP, the large abundance of gluons in this phase provides a possible channel via gluon fusion. This channel may also give rise to the production of other neutral mesons, particularly $\eta$, and maybe also $\eta'$, depending on the energy of the gluon pair. In HIC, due to their larger vertex coupling factor, gluons are expected to display a higher energy loss in the medium than quarks. These differences may introduce a change in the suppression patterns observed for $\pi^0$ and $\eta$ mesons~\cite{Acharya_2018}, due to the comparatively larger gluon component in the $\eta$ meson (a two-gluon component in the later). 
Gluon fusion into a single $\pi^0$ (or $\eta$) has been analyzed long ago in the vacuum \cite{PhysRevD.19.2188,SHIFMAN1991341}.
The fusion into a $\pi^0$ in the vacuum
provides a probe of isospin symmetry breaking, being therefore quite suppressed. On the other hand, the fusion into $\eta$ in the vacuum is a probe of flavor symmetry breaking (FSB), which is considerably larger than the pion production case. In the present work, we investigate the gluon fusion process, leading to neutral pion $\pi^0$, $\eta$, or $\eta'$ production, respectively, in the presence of a constant and uniform magnetic field. The later constitutes a phenomenological approximation to the early conditions in the QGP where the gluon fusion occurs, where it is expected that some remnants of the magnetic field generated by the initial nuclear collision might still be present prior to thermalization. Therefore, in order to model this physical scenario, we consider (constituent) quark propagators which are dressed by the magnetic field, but in {\it{the weak field limit}}, and at zero temperature.
The three neutral pseudoscalar mesons, in the framework of the 
standard quark model, are normalized 
quark-antiquark states as follows:
\begin{eqnarray}
 | \pi^0 > &\simeq& P_3
  =
    \frac{1}{\sqrt{2}}
    ( | u \bar{u} >  
  - | d \bar{d} > ),
 \;\;\;\;
 | P_8 > = \frac{1}{\sqrt{6}}
 ( |u\bar{u} > 
 + | d \bar{d} > - 2 | s \bar{s} >
 ),
 \;\;\;\;
 | P_0 > = \frac{1}{\sqrt{3}}
 ( |u\bar{u} > 
 + | d \bar{d} > + | s \bar{s} >
 ).
\end{eqnarray}
It is well known that chiral anomaly and flavor symmetry breaking
provide a $\eta-\eta'$, and $\pi^0$, mixing \cite{Feldmann,Kroll}.
The neutral pion mixing is very small and will be neglected in this work.
The effect of this mixing on the gluon fusion to the $\eta$ and the $\eta'$ will also be addressed.
It will be assumed that  gluons couple to constituent (dressed) 
quarks,  with  large quark/antiquark
 effective masses,
producing a neutral pion, $\eta$, or an $\eta'$, along the standard framework.
Whereas  mesons are  non-fundamental particles observed in 
the confined phase, and usually described by effective
models or effective field theories (EFT), gluons are expected to move 
in the deconfined phase.
Therefore, along this work it is assumed a sudden hadronization
and both the
contribution of quark- fragmentation and the location of these processes will not be addressed.
We will make use of the 
Weinberg's large Nc EFT 
\cite{Weinberg-2010} that was derived in Refs.
\cite{Braghin2016a,Braghin2018c,Braghin2019} in the vacuum and under weak magnetic field.
This EFT
 describes interactions of 
constituent quarks, gluons and pseudo-Goldstone bosons
being defined in the non-linear realization of 
chiral symmetry with axial and vector pion couplings to 
constituent quarks.
However, the linear pion coupling can be obtained by
a pion field redefinition leading to 
the linear realization of chiral symmetry.
This EFT has been already applied to the calculation 
of form factors in the vacuum and under magnetic fields for example in Refs \cite{Braghin2018,Braghin2019,Braghin2020a,PhysRevD.109.034014}.

 This work is organized as follows.
 In the section~\ref{GF_vacuum},
 the two gluon fusion into a single pseudoscalar neutral meson 
 ($\pi^0, \eta$ or $\eta'$) will be 
 calculated in the vacuum by 
 considering the corresponding 
 triangle diagram.
 Two different kinematical situation will be discussed, to be referred as Case~(a) and Case~(b), respectively.
In section \ref{GF_B}
the same diagrams in a weak magnetic field  will be calculated, by means of the leading contribution that 
comes from the (constituent) quark 
propagator. Other contributions should be sub-leading  as discussed below.
In Appendix~\ref{App:kinematics}, the kinematical variables are discussed in details, with the corresponding cross-section presented in Appendix~\ref{App:Cross_section}. In Appendix~\ref{App:Integrals}, we present the analytical expressions for the phase-space integrals involved in the Feynman diagrams computed for the probability amplitudes. 
Numerical results are presented in 
section \ref{numerics}, and conclusions drawn in section \ref{conclusion}.

\section{ Gluon fusion diagrams}

As discussed in the Introduction, gluon fusion will be
 analyzed by means of its leading process, that is 
one single quark loop, that must be a triangle diagram.
This diagram may be considered an application of the 
Weinberg's Large Nc EFT,
 being the pseudoscalar pion coupling to 
constituent quark obtained by means of a pion field redefinition.
Besides the gluon-quark coupling, that will be replaced by a
gluon-constituent quark coupling, along the lines of \cite{Weinberg-2010}, the pseudoscalar meson, $P_i$ (for $i=0,...8$), coupling to quarks
is given by
\begin{eqnarray}
 {\cal L}_{\pi-q} &=&
 G_{ps} P_i \; 
( \;  \bar{\psi} i \gamma_5 \lambda_i
\psi \; ),
\end{eqnarray}
where $i=0,...N_f^2-1$, and, by simply considering that constituent quarks and 
the nucleon couples in the same way: $G_{ps} \simeq 13-13.8$
from phenomenological investigations~\cite{GPS}, in agreement with the Goldberger Treiman relation (GTR)~\cite{GTR}.

\subsection{ Gluon fusion in the vacuum}
 \label{GF_vacuum}

\begin{figure}[ht!]
\centering
\includegraphics[width=90mm]{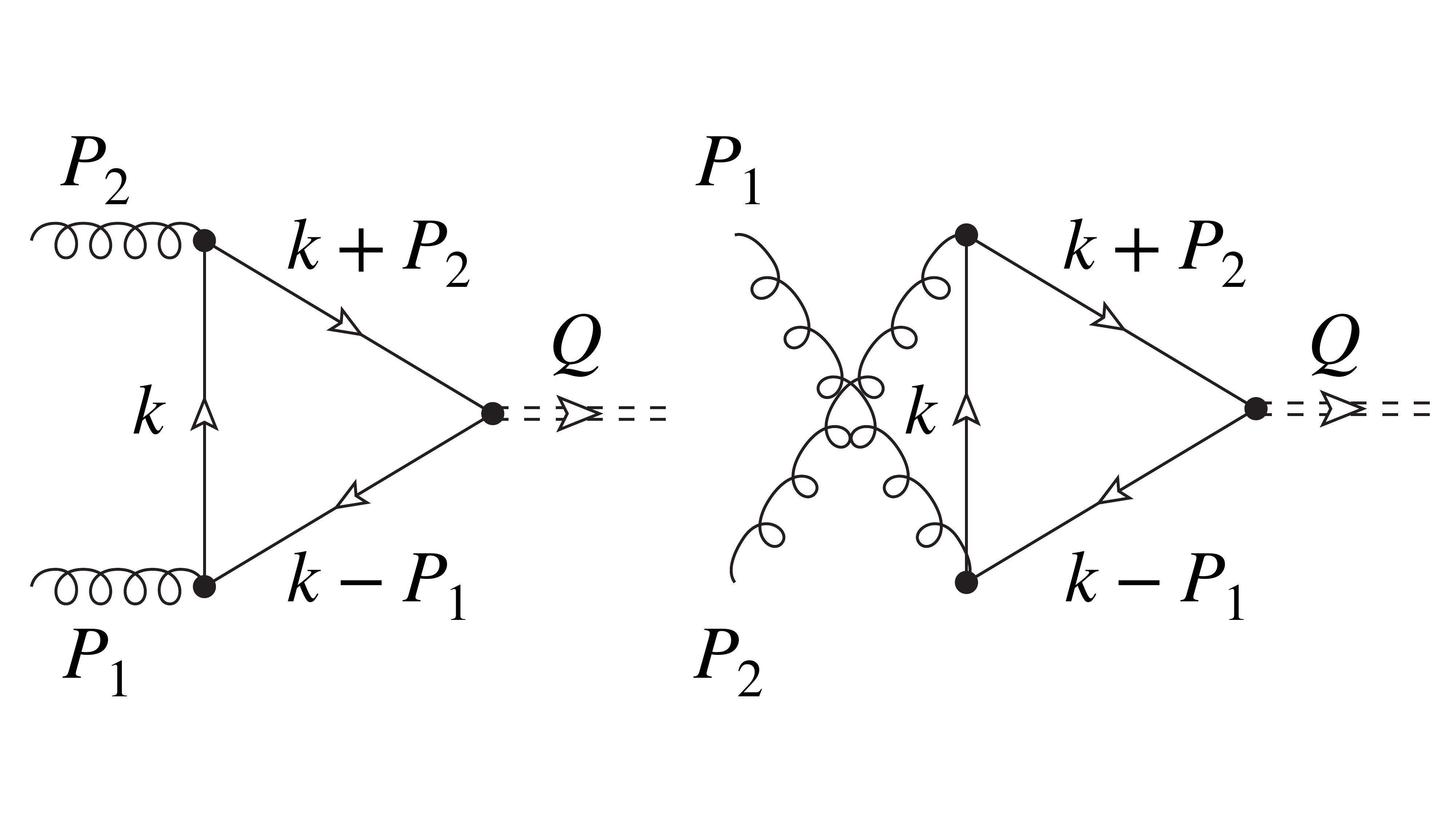}
 \caption{ 
\small
Triangle diagram for the gluon (curly line) fusion into a single PS meson (double dashed line).
}\label{fig:fig1}
\end{figure}

By considering the pseudoscalar meson 
($\pi^0$, $\eta$, or $\eta'$) coupling to quarks, and gluons with momenta 
$P_1$ and $P_2$,
the diagrams presented in Fig. 
\eqref{fig:fig1} are  calculated, for a specific
quark flavor $f=u,d,s$, as
\begin{eqnarray}
I_{gl.fusion}^{(f)} &=&
G_{ps} g^2 Tr_{D,C}
\left[
\frac{\gamma^\mu \lambda^a}{2}
S_0 (k+P_2) i \gamma^5 \lambda^i
S_0 (k-P_1) \frac{ \gamma^\nu  \lambda^b}{2}
S_0(k) \right]
\epsilon_\mu (P_1) \epsilon_\nu (P_2)
\end{eqnarray}
where $g \sim 1$ is the quark-constituent gluon coupling constant (taken as constant),
 $a,b$ are color indices of the adjoint representation, with corresponding Gell Mann matrices $\lambda^a$,  
and $i,j$ are   flavor  indices of the adjoint representation, in which $i=3$, $8$, and $0$ label the neutral pion $\pi^0$, the $\eta$,
and the $\eta'$, respectively.
The quark propagator, for each flavor species,
is written in terms of a quark constituent mass $M$, such that 
$S_0 (k) = i (\slashed{k} - M+ i\epsilon )^{-1}$, 
and $\epsilon^\mu(P)$ are the 
gluon polarizations.

After some standard steps to introduce 
Feynman parameters to integrate the internal quark momenta,
the amplitude can be written as:
\begin{eqnarray}
I_{gl.fusion}^{(f)} &=&
\frac{i 2 \delta_{ab} G_{ps} N_c M}{(4\pi)^2}
\int_{xy} \frac{ \epsilon_\mu(P_1) \epsilon_\nu(P_2)}{
M^2_f - 2 x y P_1 \cdot P_2 }
\nonumber
\\
&\times&
\left[
\epsilon^{\mu\alpha\beta\nu} (P_1(1-x) + P_2y)^\alpha
( P_2 (y-1) - P_1x)^\beta + 
\epsilon^{\mu\alpha\nu\delta}
( P_1 (1-x) + P_2 y)^\alpha (P_2y - P_1x)^\delta
\right.
\nonumber
\\
&-& \left.
\epsilon^{\mu\beta\nu\gamma} (P_2 (y-1) - P_1 x)^\beta
(P_2y - P_1 x)^\gamma
\right].
\end{eqnarray}
The resulting integral is UV finite, just like other triangle diagrams, 
such as for the anomalous 
neutral pseudoscalar meson decay into two photons.
The resulting integrals have the following form:
\begin{eqnarray}
\mathcal{K}_n^{(q)} &=& \int_0^{1} dy\, y^q I_{n}^{(1)}(y),
\nonumber
\\
I_n^{(1)} &=&
\int_0^{1-y} d x \frac{ x^n}{ ( \alpha x + M^2)}, \;\;\;\;\;\;
\alpha = - 2 y P_1 \cdot P_2.
\end{eqnarray}
The resulting integrals are shown in the Appendix~\ref{App:Integrals}.
The coefficient for each single quark species will be given by
\begin{eqnarray}
C_{ggp}^{vac} = 
\frac{ i  2 \delta_{ab} N_c M G_{ps}}{(4\pi^2)}.
\end{eqnarray}

There are many possible choices for the
kinematics of two gluons participating in the  fusion  process.
The two gluons (and their momenta) will be labeled as gluon 1 and gluon 2, and two very different kinematical configurations, referred to as case (a) and case (b), respectively, are discussed below.

\subsubsection{ Case (a)}
\label{sec:case-a}
Let us consider one gluon, 3-momentum $\mathbf{P}_1$, parallel to the magnetic field $\mathbf{B}$ and the second
gluon, with $\mathbf{P}_2$, has  arbitrary direction.
These momenta can be parameterized as
\begin{eqnarray}
P_1^{\mu} &=& (E_1, 0, 0, -|\mathbf{P}_1|)\nonumber\\
P_2^{\mu} &=& (E_2, |\mathbf{P}_2|\sin\theta\cdot\cos\varphi,|\mathbf{P}_2|\sin\theta\cdot\sin\varphi,|\mathbf{P}_2|\cos\theta)
\end{eqnarray}
The corresponding mutually orthogonal polarization vectors (using the notation $\epsilon_j^{a,\mu}$, with $j = 1,2$ the two independent directions and $a = 1, 2$ the two gluons), are
\begin{eqnarray}
\epsilon_1^{1,\mu} &=& (0,1,0,0), \;\;\;\;\;
\epsilon_1^{2,\mu} =(0,\cos\theta\cdot\cos\varphi,\cos\theta\cdot\sin\varphi,-\sin\theta)\nonumber\\\nonumber\\
\epsilon_{2}^{1,\mu} &=& (0,0,1,0), 
\;\;\;\;\;
\epsilon_2^{2,\mu} = (0,-\sin\varphi,\cos\varphi,0).\nonumber\\
\end{eqnarray} 
For massless on-shell gluons with energies $E_1$ and $E_2$,
the resulting gluon fusion amplitude into a single quark flavor $f$ will be given by

\begin{eqnarray}
T_{ggp,f}^{vac} &=& 
- C_{ggp}^{vac}
\mathcal{K}_0^{(0)}
E_1 E_2
\left\{
[   c_\theta  +  1 ]
( 1 + c_\varphi  + c_\theta c_\varphi (1 + c_\varphi) 
+
[   c_\theta +  1  ]
c_\theta s_\varphi^2
+
[   s_\theta s_\varphi ]
  s_\varphi^2 
+
[ s_\theta c_\varphi ]
s_\varphi ( 1  +c_\varphi)
\right\},
\end{eqnarray}
 where we defined, for notational convenience, $c_{\theta}\equiv \cos\theta$, $c_{\varphi}\equiv \cos\varphi$, $s_{\theta}\equiv\sin\theta$, and $s_{\varphi} \equiv \sin\varphi$, respectively.

If we average over the azimuthal angle $\varphi$, we obtain
\begin{eqnarray} 
\langle T_{ggp,f}^{vac} \rangle
&=& - C_{ggp}^{vac}
\mathcal{K}_0^{(0)}
E_1 E_2
\left\{
\left(   c_\theta  +  1 \right)
\left( 1  + c_\theta  
+
\frac{1}{2} c_\theta^2\right) 
\right\}.
\end{eqnarray}

Although this amplitude seems
 explicitly proportional to the energy of the gluons, there is 
an inverse behavior in the integral $\mathcal{K}_0^{(0)}$ that compensates this effect.

For the kinematics, we notice that energy-momentum balance of the process indicates that, for $Q$ the pion 4-momentum
\begin{equation}
Q = P_1 + P_2 \Longrightarrow Q^2 = (P_1 + P_2)^2 = 2 P_1\cdot P_2
\end{equation}
where we used that on-shell gluons satisfy $P_1^2 = P_2^2 = 0$ ($|\mathbf{P}_1|=E_1$, and $|\mathbf{P}_2| = E_2$, respectively). Moreover, in terms of the pion invariant mass $Q^2 = s$, we have
\begin{equation}
P_1\cdot P_2 = E_1 E_2 -\mathbf{P}_1\cdot\mathbf{P}_2 = E_1 E_2\left( 1 + \cos\theta \right)  \equiv \frac{1}{2}s > 0.
\label{eq_P1P2_a}
\end{equation}

All the magnitudes are defined in therms of the three independent variables $s$, $p_T$ and rapidity $y$, as follows (see Appendix~\ref{App:kinematics} for details)
\begin{eqnarray}
E_1 &=& \frac{s}{2\left( p_T + \sqrt{\frac{p_T^2 + s}{1 - \tanh^2(y)}} \right)},\nonumber\\
E_2 &=& \sqrt{\frac{p_T^2 + s}{1 - \tanh^2(y)}} - \frac{s}{2\left( p_T + \sqrt{\frac{p_T^2 + s}{1 - \tanh^2(y)}} \right)},
\label{eq_E2}
\end{eqnarray}
with
\begin{eqnarray}
\tan\theta = \frac{ \sqrt{\frac{p_T^2 + s}{1 - \tanh^2(y)}}  }{ p_T + \frac{s}{2\left( p_T + \sqrt{\frac{p_T^2 + s}{1 - \tanh^2(y)}} \right)} } \tanh(y).
\label{eq_tanycasea}
\end{eqnarray}

\subsubsection{ Case (b)}
\label{sec:case-b}
In this case, 
one gluon, with 3-momentum  $\mathbf{P}_1$, is orthogonal to the magnetic field $\mathbf{B}$, and the second gluon, with  $\mathbf{P}_2$, has  arbitrary direction. 
This can be parameterized as follows
\begin{eqnarray}
P_1^\mu &=& (E_1,  |\mathbf{P}_1| \cos \phi,  |\mathbf{P}_1| \sin \phi, 0),\nonumber\\
P_2^\mu &=& (E_2, |\mathbf{P}_2| \sin \theta . \cos \varphi , 
 |\mathbf{P}_2| \sin \theta . \sin \varphi,   |\mathbf{P}_2| \cos \theta).
\end{eqnarray}
The corresponding mutually orthogonal polarization vectors are
\begin{eqnarray}
\epsilon_1^{1,\mu} &=& (0,0,0,1), \,\,\,\,\,
\epsilon_1^{2,\mu} = (0,\cos\theta\cdot\cos\varphi,\cos\theta\cdot\sin\varphi,-\sin\theta),
\nonumber\\
\epsilon_{2}^{1,\mu} &=& (0,\sin\phi,-\cos\phi,0), 
\,\,\,\,\,
\epsilon_2^{2,\mu} = (0,-\sin\varphi,\cos\varphi,0).\nonumber\\
\end{eqnarray}
For on-shell massless gluons with the kinematics of case~(b),
the amplitude to produce a single $f$ quark is given by:
\begin{eqnarray}
T_{ggp,f}^{vac} &=& 
C_{ggp}^{vac}
\mathcal{K}_0^{(0)} E_1 E_2    \left\{ (    s_\theta s_\varphi  - s_\phi  ) 
[
-  (c_\theta c_\varphi - s_\phi)
-
 s_\phi^2 ] 
+ (
s_\theta c_\varphi-  c_\phi)  
[
 ( c_\theta s_\varphi + c_\phi)
-
  c_\phi s_\theta
  ] \right.
  \nonumber\\
  && \left. +
( c_\theta )   
[ -
 s_\phi (c_\theta s_\varphi + c_\phi)
-  
 c_\phi (c_\theta c_\varphi - s_\phi) ] 
 \right\}.
\end{eqnarray}

If we average over the independent azimuthal angles $\varphi$ and $\phi$, we obtain
\begin{eqnarray}
\langle T_{ggp,f}^{vac} \rangle &=& 
C_{ggp}^{vac}
\mathcal{K}_0^{(0)} E_1 E_2    \left\{ \frac{s_{\theta}}{2} -1 \right\}
.
\end{eqnarray}

As shown in detail in Appendix~\ref{App:kinematics}, all the magnitudes are defined in therms of the three independent variables $s$, $P_T$, and 
rapidity $y$. We choose here particular values for $\phi$ and $\varphi$, such that
\begin{equation}
\mathcal{C} = \cos(\varphi-\phi) = \left\{\begin{array}{cc} +1, & \phi = \varphi = 0\\-1, & \phi = 0,\,\varphi = \pi. \end{array}\right.
\end{equation}

As shown in Appendix~\ref{App:kinematics}, under the condition $\mathcal{C}=\pm 1$ stated above, we obtain the explicit expressions
\begin{eqnarray}
E_1 &=& \sqrt{\frac{1 \pm \tanh(y)}{1 \mp \tanh(y)}}\frac{s}{2\sqrt{s + p_T^2}},\\
E_2 &=& \frac{2 p_T^2 + s\left( 1 \mp \tanh(y) \right)}{2\sqrt{p_T^2 + s}\sqrt{1 - \tanh^2(y)} },
\label{eq_ECaseb}
\end{eqnarray}
with
\begin{eqnarray}
\mathcal{C}\tan\theta = \frac{2 E_1 E_2 - s}{2 p_T E_1}.
\label{eq_tanycaseb}
\end{eqnarray}

\subsection{ Gluon fusion in a relatively weak magnetic field}
\label{GF_B}

\begin{figure}[ht!]
\centering
\includegraphics[width=150mm]{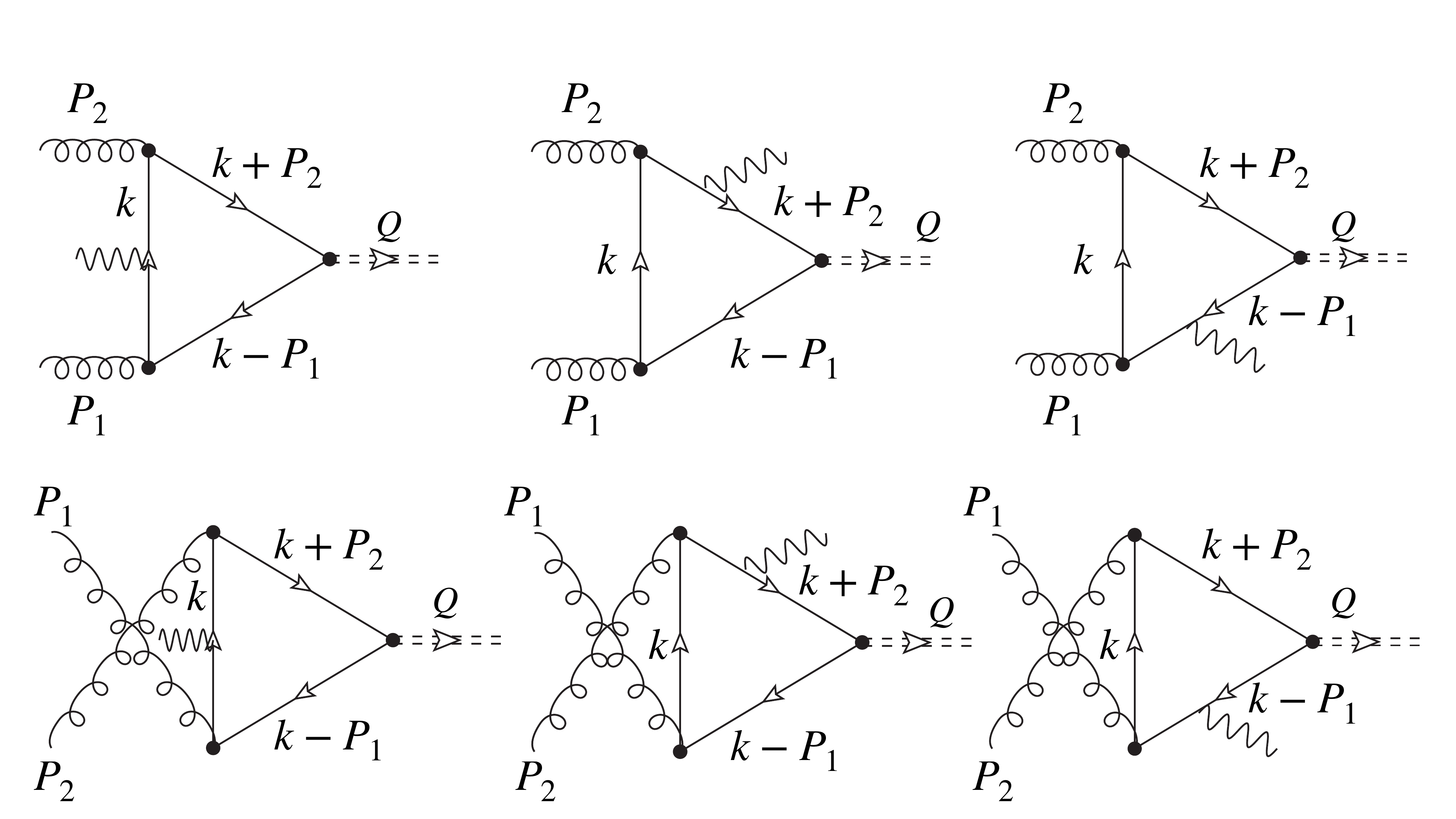}
 \caption{ 
\small
Feynman Diagrams for weak magnetic field: The magnetic field insertion is represented by a small wiggly line in the quark propagators. Gluons correspond to the curly lines.
}
\label{fig:fig2}
\end{figure}

Magnetic field corrections to the gluon fusion into a single pseudoscalar 
meson will be considered by means of  two (leading) effects.
Firstly, for magnetic fields relatively weak with respect to the 
quark (constituent) mass (i.e. $|eB| \ll M^2$) the quark propagator with the leading correction can be written as
\cite{Chyi2000}
\begin{eqnarray}  \label{quark-prop}
S (k) &=&  S^{(0)} (k) + S^{(1)} (k) (e B) = i 
\frac{ \slashed{k} + {M} }{ k^2 - {M}^2 } 
- \gamma_1 \gamma_2
 \frac{
 {M}^2(  \gamma_0 k^0 - \gamma_3 k^3  +  {M}  )
 }{
(k^2 - {M}^2)^2 }  \hat{Q} \frac{e B
}{ {M}^2},
\end{eqnarray}
where $\hat{Q}$ is the quark charge matrix.
The corresponding Feynman diagrams are depicted in 
Fig.~\eqref{fig:fig2}.

In this weak magnetic field approach, all the
leading terms 
linear in $eB$ are collected for each quark internal line separately,
so that there are three contributions for each diagram, as shown in Fig.~\eqref{fig:fig2}, where the (single) magnetic field insertion is depicted as a wiggly line on the fermion propagator.
 Since the quark loop involves 
one single quark flavor, with an unique quark charge,
the Schwinger phase is trivially zero.

As a second effect due to the external  magnetic field, 
quark (constituent) masses should be modified as obtained
from gap equations~\cite{ADHIKARI_2026,NJL-2022}.
This is of order ${\cal O} (eB)$.
We could also consider other effects that can be expected to be of 
higher order in 
 $(eB)^n$, for $n \geq 2$, and therefore smaller, such as the 
behavior of the pseudoscalar meson coupling to quarks 
\cite{Braghin2020a} and the effect on (neutral) external 
lines. These will not be considered in the present work.

The diagrams of Fig.~\eqref{fig:fig2}-a, for 
incoming photon polarizations 
$(\epsilon^\mu)(P_1)$ and
$(\epsilon^\nu) (P_2)$, and for a single quark species $f$,
can be written as
\begin{eqnarray}
&&  Tr_{D,C} \left(
\gamma_\mu \frac{\lambda_a}{2}
S_f (k + P_2)
\left[ i\gamma_5 \lambda_i ; Q^\rho \gamma_\rho \gamma_5 \lambda_i \right]
S_f ( k  - P_1) \gamma_\nu \frac{\lambda_b}{2}
S_f ( k)
\right)   (\epsilon^\mu)  (P_1) (\epsilon^\nu)  (P_2),
\end{eqnarray}
whereas diagrams in Fig.~\eqref{fig:fig2}-b are obtained, roughly speaking, by exchanging 
$(\epsilon^\mu)(P_1) \leftrightarrow (\epsilon^\nu) (P_2)$.

The final expressions for these diagrams will be analyzed below
separately for the same kinematical configurations defined before as case (a) (Sec.~\ref{sec:case-a}), and case (b) (Sec~\ref{sec:case-b}), respectively.

\subsubsection{ Case (a)}

Here, we shall consider the same kinematical configuration as in Sec.~\ref{sec:case-a}. After standard calculations by using the Feynman trick and 
parameters,
the final amplitude for each quark flavor 
can be written as:
\begin{eqnarray}
\label{ggp-a-B-a}
T_{ggp,f}^{(B)}  &=& C_{ggp}^{(B)} 
K 
\left\{ \left(   
- \mathcal{J}_0^{(3)} + 3\mathcal{J}_0^{(2)}  - 3 \mathcal{J}_0^{(1)} + 
\mathcal{J}_0^{(0)}  \right)
\left[    (c_\theta c_\varphi - s_\varphi )
+ (
c_\theta s_\varphi + c_\varphi 
)
\right]
-
2 \mathcal{J}_1^{(1)}  
\left[ -  c_\theta  c_\varphi + s_\varphi  - c_\theta s_\varphi - c_\varphi 
\right] 
\right.
\nonumber
\\
 &&  \left. 
- \mathcal{J}_1^{(1)}
\left[ 
- c_\theta c_\varphi  + c_\theta s_\varphi + c_\varphi
+ s_\varphi
\right] 
+
  \left(  
 2 \mathcal{J}_1^{(1)} - \mathcal{J}_1^{(0)}
\right)
\left[  (c_\theta c_\varphi - s_\varphi )
+   (
c_\theta s_\varphi + c_\varphi 
)
\right] \right.
\nonumber
\\
&& \left.  
- \mathcal{J}_2^{(0)}
\left[ -  c_\theta  c_\varphi + s_\varphi  - c_\theta s_\varphi - c_\varphi 
\right]  
+
  \left[ 
\mathcal{J}_2^{(0)}
- 2 
\mathcal{J}_2^{(1)}
\right] 
\left[   (c_\theta c_\varphi - s_\varphi )
+
  (
c_\theta s_\varphi + c_\varphi 
)
\right]
\right\}
\nonumber\\
&&  
+ C_{ggp}^{(B)}\left\{
2 ( E_1 P_2^x - E_2 P_1^x)  
 \left( {\cal I}_0^0  - 2 {\cal I}_1^1 +2 {\cal I}_1^0 \right)
 ( s_\varphi)  
-
 2 ( E_1 P_2^y - E_2 P_1^y )s_{\varphi}
 \left( {\cal I}_0^0  - 2 {\cal I}_1^1 +2 {\cal I}_1^0 - 3  {\cal I}_2^1 + 3 {\cal I}_0^2 \right)\right\}.
\end{eqnarray}
Here, we defined the magnetic field dependent coefficients
\begin{eqnarray}
C_{ggp}^{(B)} = N_c \frac{ eB}{M^2}
\times \frac{ M^3 G_{ps}}{(4\pi)^2}
=   \frac{C_{ggp}^{vac} M^2}{2}
\frac{ eB}{M^2},
\end{eqnarray}
where,
for on-shell massless gluons,
we have 
\begin{eqnarray}
K &\equiv& P_2^0 P_1^3 - P_2^3 P_1^0
=
 - E_1 E_2 \; 
 ( 1 + c_\theta),
\end{eqnarray}
\begin{eqnarray}
E_1 P_2^x - E_2 P_1^x = E_1 E_2 s_{\theta} c_{\varphi}, 
\end{eqnarray}
and
\begin{eqnarray}
E_1 P_2^y - E_2 P_1^y = E_1 E_2  s_{\theta}s_{\varphi}. 
\end{eqnarray}

The corresponding probability amplitude is then given by the expression
\begin{eqnarray}
\label{ggp-a-B}
T_{ggp,f}^{(B)}  &=& C_{ggp}^{(B)} 
K 
\left\{ \left(   
- \mathcal{J}_0^{(3)} + 3\mathcal{J}_0^{(2)}  - 3 \mathcal{J}_0^{(1)} + 
\mathcal{J}_0^{(0)}  \right)
\left[    (c_\theta c_\varphi - s_\varphi )
+ (
c_\theta s_\varphi + c_\varphi 
)
\right]
-
2 \mathcal{J}_1^{(1)}  
\left[ -  c_\theta  c_\varphi + s_\varphi  - c_\theta s_\varphi - c_\varphi 
\right] 
\right.
\nonumber
\\
 &&  \left. 
- \mathcal{J}_1^{(1)}
\left[ 
- c_\theta c_\varphi  + c_\theta s_\varphi + c_\varphi
+ s_\varphi
\right] 
+
  \left(  
 2 \mathcal{J}_1^{(1)} - \mathcal{J}_1^{(0)}
\right)
\left[  (c_\theta c_\varphi - s_\varphi )
+   (
c_\theta s_\varphi + c_\varphi 
)
\right] \right.
\nonumber
\\
&& \left.  
- \mathcal{J}_2^{(0)}
\left[ -  c_\theta  c_\varphi + s_\varphi  - c_\theta s_\varphi - c_\varphi 
\right]  
+
  \left[ 
\mathcal{J}_2^{(0)}
- 2 
\mathcal{J}_2^{(1)}
\right] 
\left[   (c_\theta c_\varphi - s_\varphi )
+
  (
c_\theta s_\varphi + c_\varphi 
)
\right]
\right\}
\nonumber
\\
&&  
+ C_{ggp}^{(B)}\left\{
2 E_1 E_2 s_{\theta}  
 \left( {\cal I}_0^0  - 2 {\cal I}_1^1 +2 {\cal I}_1^0 \right)
 ( s_\varphi)  
-
 2 E_1 E_2  s_{\theta}s_{\varphi}^2
 \left( {\cal I}_0^0  - 2 {\cal I}_1^1 +2 {\cal I}_1^0    
  - 3  {\cal I}_2^1 + 3 {\cal I}_0^2 \right)\right\}.
\end{eqnarray}

If we average over the azimuthal angle $\varphi$, we obtain
\begin{eqnarray}
\label{ggp-a-B-avg}
\langle T_{ggp,f}^{(B)}\rangle 
&=& C_{ggp}^{(B)} 
\left\{ 
-
 E_1 E_2\,s_{\theta}
 \left( {\cal I}_0^0  - 2 {\cal I}_1^1 +2 {\cal I}_1^0  - 3  {\cal I}_2^1 + 3 {\cal I}_0^2\right) \right\}.
\end{eqnarray}

The remaining integrals in Feynman parameters are the following
\begin{eqnarray}
&& \mathcal{J}_n^{(q)} = \int_0^1 dy\, y^{q}\,I_n^{(2)}(y),
\nonumber
\\
\mbox{where} \;\;\;\;\; &&
I_n^{(2)}(y;\alpha) \rightarrow \int_0^{1-y}dx \frac{x^n}{\left[ \alpha x + M^2\right]^2}.
\end{eqnarray}
The  explicit analytical expressions for the 
integrals needed in Eq.~\eqref{ggp-a-B}
 are listed in the Appendix~\ref{App:Integrals}.

\subsubsection{ Case (b)}

Here, we shall consider the same kinematical configuration as in Sec.~\ref{sec:case-b}. After standard steps, by introducing Feynman parameters, the amplitude can be written in terms of the same integrals
discussed above for case (a), whose analytical expressions are given in Appendix~\ref{App:Integrals}.
The final amplitude for each quark flavor for case (b), 
after the contraction with polarizations and summations,
 is given by
\begin{eqnarray}
T_{ggp,f}^{(B)}  &=& C_{ggp}^{(B)} 
K \left\{ \left(    (\mathcal{J}_0^{(0)}- 2 \mathcal{J}_0^{(1)} + \mathcal{J}_0^{(2)})
   \right)
 \left[   ( s_\phi c_\theta c_\varphi
- s_\phi^2 )
+
( - c_\phi c_\theta s_\varphi -  c_\phi^2
)  - s_\theta \right]
 \right.
\nonumber
\\
&& \left.
-
2 
 \mathcal{J}_1^{(1)} 
\left( 
s_\theta^2 + c_\phi^2 - s_\phi c_\theta c_\varphi
+ c_\phi c_\theta s_\varphi 
+s_\theta
\right)
  - 
\mathcal{J}_1^{(1)}
\left[ 
  ( - s_\phi^2 + s_\phi c_\theta c_\varphi )  
+
 ( 
-c_\phi^2 
- c_\phi c_\theta s_\varphi
 ) - 
 s_\theta \right] \right.
\nonumber
\\
  &&  \left.
+  \left(  
 2 \mathcal{J}_1^{(1)}
\right)
  \left[   ( s_\phi c_\theta c_\varphi
- s_\phi^2 )
+
( - c_\phi c_\theta s_\varphi -  c_\phi^2
) - s_\theta \right] \right.
\nonumber
\\
&& \left.
-  \mathcal{J}_2^{(0)}
\left( 
s_\theta^2 + c_\phi^2 - s_\phi c_\theta c_\varphi
+ c_\phi c_\theta s_\varphi 
+ s_\theta
\right)
+
  \left[ 
\mathcal{J}_2^{(0)} - 2
\mathcal{J}_2^{(1)}
\right] 
 \left[   ( s_\phi c_\theta c_\varphi
- s_\phi^2 )
+
( - c_\phi c_\theta s_\varphi -  c_\phi^2
) - s_\theta  \right]
\right\}
\nonumber
\\
&+& C_{ggp}^{(B)} 
\left\{ 
   (\mathcal{J}_0^{(0)}
   + \mathcal{J}_0^{(2)}
   - 2 \mathcal{J}_0^{(1)}
   )
[ E_1 P_2^x - E_2 P_1^x]
  \left[   s_\theta s_\phi 
- s_\theta c_\phi  
 \right]\right.
\nonumber
\\
    &&  
   \left. +
\left[ 
[ E_1 P_2^x - E_2 P_1^x  ]
 s_\theta s_\phi 
-  
[ E_1 P_2^y - E_2 P_1^y ]
 s_\theta c_\phi
\right]
(
\mathcal{J}_0^{(0)}
-  3\mathcal{J}_0^{(1)}
+ 3 \mathcal{J}_0^{(2)}
- \mathcal{J}_0^{(3)}
)
\right.
\nonumber
\\
&&  
\left. 
+
\left[
[ E_1 P_2^x - E_2 P_1^x  ] s_\theta s_\phi 
- 
[ E_1 P_2^y - E_2 P_1^y ]   s_\theta c_\phi 
\right]
\left( - 3 \mathcal{J}_1^{(0)}
+ 3 \mathcal{J}_1^{(1)} \right) 
\right.
\nonumber
\\
&&   \left.
- 
\left[ [ E_1 P_2^x - E_2 P_1^x  ] s_\theta s_\phi 
- [ E_1 P_2^y - E_2 P_1^y ]
s_\theta c_\phi  \right]
\left(
\mathcal{J}_2^{(0)}
- 2 \mathcal{J}_2^{(1)}
\right)
 \right\},
\end{eqnarray}
where
\begin{eqnarray}
K &\equiv& P_2^0 P_1^3 - P_2^3 P_1^0
 = - E_1 E_2 c_\theta,
 \\
C_{ggp}^{vac} &=& 
\frac{ i   \delta_{ab} N_c M G_{ps}}{(4\pi^2)},
\end{eqnarray}
\begin{eqnarray}
E_1 P_2^x - E_2 P_1^x = E_1 E_2 \left( s_{\theta} c_{\varphi} - c_{\phi} \right),
\end{eqnarray}
and
\begin{eqnarray}
E_1 P_2^y - E_2 P_1^y = E_1 E_2 \left( s_{\theta}s_{\varphi} - s_{\phi} \right).
\end{eqnarray}
If we average over the azimuthal angles $\varphi$ and $\phi$, we obtain
\begin{eqnarray}
\langle T_{ggp,f}^{(B)} \rangle &=& C_{ggp}^{(B)} 
K \left\{ \left(    (\mathcal{J}_0^{(0)}- 2 \mathcal{J}_0^{(1)} + \mathcal{J}_0^{(2)})
   \right)
 \left[   -1 - s_\theta \right]
-
2 
 \mathcal{J}_1^{(1)} 
\left( 
s_\theta^2 + 1/2 
+s_\theta
\right)
  - 
\mathcal{J}_1^{(1)}
\left[ 
  -1 - 
 s_\theta \right] \right.
\nonumber
\\
  &&  \left.
+   
 2 \mathcal{J}_1^{(1)}
  \left[   -1 - s_\theta \right] 
-  \mathcal{J}_2^{(0)}
\left( 
s_\theta^2 + 1/2  
+ s_\theta
\right)
+
  \left[ 
\mathcal{J}_2^{(0)} - 2
\mathcal{J}_2^{(1)}
\right] 
 \left[   -1 - s_\theta  \right]
\right\}
\nonumber\\
&& +
   C_{ggp}^{(B)} (\mathcal{J}_0^{(0)}
   + \mathcal{J}_0^{(2)}
   - 2 \mathcal{J}_0^{(1)}
   )
\frac{E_1 E_2}{2} s_{\theta}.
\end{eqnarray}

\subsection{ Gluon fusion rates}
 
The probabilities to form  the
neutral pion, and the states $P_8$ and $P_0$,
that lead to the 
$\eta$ and $\eta'$, for each kinematical case,
$\alpha = a,b$, and for  fixed
 gluon polarizations, can be written as:
\begin{eqnarray}
 T^{gg\rightarrow P_3,vac}_{\alpha} 
&=& 
\frac{1}{\sqrt{2}}
\left(
T^{vac,\alpha}_{ggp,u} 
- T^{vac,\alpha}_{ggp,d}\right) ,
\\
  T^{gg\rightarrow P_8,vac}_{\alpha}  
&=& 
\frac{1}{\sqrt{6}}
\left(
T^{vac,\alpha}_{ggp,u} 
+ 
T^{vac,\alpha}_{ggp,d}
- 2  T^{vac,\alpha}_{ggp,s}\right) ,
\\
 T^{gg\rightarrow P_0,vac}_{\alpha} 
&=& 
\frac{1}{\sqrt{3}}  
\left(
T^{vac,\alpha}_{ggp,u} 
+ 
T^{vac,\alpha}_{ggp,d}
+  T^{vac,\alpha}_{ggp,s}\right).
\end{eqnarray}
Note that the probability density for the 
gluon fusion into the $\pi^0$ or the $\eta$, 
but not the $\eta'$, in the vacuum becomes proportional to the quark mass differences, 
$(M_{f_1} - M_{f_2})^n$
(for $n=1,2,..$), 
as already noted in \cite{PhysRevD.19.2188,SHIFMAN1991341}, and hence it must disappear in the chiral limit.
This effect reduces the resulting probabilities, with a stronger reduction in the pion case, since ISB is smaller than FSB.
For the $\eta'$, by analyzing the $P_0$  flavor state,  however, the resulting probability density is not reduced.

The  corresponding equations for the process under a   magnetic field
can be written as (for cases $\alpha = a,b$):
\begin{eqnarray}
\label{mesons-Bfield}
   T^{gg\rightarrow P_3,B}_{\alpha}  
&=& 
  T^{gg\rightarrow P_3,vac}_{\alpha}  + 
\frac{  2  T^{(B),\alpha}_{ggp,u} 
+ T^{(B),\alpha}_{ggp,d} }{18}
\nonumber
\\
  T^{gg\rightarrow P_8,B}_{\alpha}  
&=& 
 T^{gg\rightarrow P_8,vac}_{\alpha}
 + 
\frac{1}{6}
\frac{  2 T^{(B),\alpha}_{ggp,u} 
- 
T^{(B),\alpha}_{ggp,d}
+ 2  T^{(B),\alpha}_{ggp,s} }{9}
\\
  T^{gg\rightarrow P_0,B}_{\alpha}  
&=& 
T^{gg\rightarrow P_0,vac}_{\alpha}  
 + 
\frac{1}{3} \frac{  2 T^{(B),\alpha}_{ggp,u} 
- 
T^{(B),\alpha}_{ggp,d}
-   T^{(B),\alpha}_{ggp,s} }{9} .
\nonumber
\end{eqnarray}

By the relative signs in the linear combinations above, it can be directly seen that the gluon fusion into a single pion exhibits a constructive interference
in the magnetic field contribution instead of destructive interference, that in contrast is present for the process in the vacuum.
For the case of the $\eta$ a similar effect occurs.
However, for the case of the $\eta'$ a destructive interference arises in the magnetic field contribution and  
$|T^{gg\rightarrow P_0,B}_{\alpha} |^2$ is suppressed.
Although the strange 
constituent quark mass is quite larger than the up and down ones, this difference is not enough to avoid the 
suppression.

\subsubsection{ Meson mixing}

Chiral anomaly and FSB lead to the well known $\eta-\eta'$ mixing
 that is parameterized with an angle ($\theta_{m}$),
\cite{Feldmann,Kroll} such that the mass eigenstates are found in terms of the flavor eigenstates given above as:
\begin{eqnarray}
\label{meson-mixing}
| \eta > &=& \cos (\theta_m) P_8  - \sin (\theta_m) P_0,
\;\;\;\;\;\;
| \eta' > =  \sin (\theta_m) P_8  + \cos (\theta_m) P_0.
\end{eqnarray}
This mixing will induce a subsequent modification of the linear combinations involved in the definition of the amplitudes $T^{(B),\alpha}_{ggp,f}$.


\section{ Numerical results}
\label{numerics}
Next, we present the values of the parameters used to produce
numerical results based on the analytical expressions presented in the previous sections.
Meson masses were taken from PDG \cite{PDG}:
\begin{eqnarray}
 m_{\pi^0} = 0.135 \; {\text{GeV}},
 \;\;\;\;
 m_{\eta} = 0.548 \; {\text{GeV}},
 \;\;\;\;
 m_{\eta'} = 0.958 \; 
 {\text{GeV}}. \;\;\;\;
\end{eqnarray}
Constituent quark masses were fixed by considering results from 
phenomenological investigations with the 
constituent quark model 
\cite{VOGL1991195,Klevansky_RevModPhys.64.649,BUBALLA2005,BIJNENS1996370,PhysRevD.77.114028,HATSUDA1994221} as follows:
\begin{eqnarray}
\label{M300}
 M_u \simeq 300\; {\text{MeV}}, 
 \;\;\;
 M_d \simeq M_u + 10\; {\text{MeV}}, 
 \;\;\;
 M_s \simeq 500\; {\text{MeV}}, \;\;\;\;\;\;\; \mbox{I}.
\end{eqnarray}
As a second set of parameters, we have also considered, for some estimations, the values
\begin{eqnarray}
\label{M330}
 M_u \simeq 330\; {\text{MeV}}, 
 \;\;\;
 M_d \simeq M_u + 10\; {\text{MeV}}. 
 \;\;\;
 M_s \simeq 500\; {\text{MeV}}, \;\;\;\;\;\;\; \mbox{II}.
\end{eqnarray}
This second set can be considered somewhat as resulting from the effect of a
weak magnetic field effect on quark masses usually extracted from phenomenology of 
NJL-model calculations \cite{ADHIKARI_2026}.

The $\eta$ and $\eta'$ couplings to nucleon are not expected to be the same as the one for the pion, mainly for the 
$\eta'$.
But this problem is full of uncertainties, and we consider the 
same value extracted from the
Goldberger Treiman relation  (GTR)~\cite{GTR} for all the three mesons. The pseudoscalar meson coupling constant to constituent quarks was associated to the pseudoscalar meson coupling to nucleon,
and taken from the GTR~\cite{GTR}
as 
\begin{eqnarray}
G_{ps} \simeq \frac{g_A M_{nuc}}{f_{ps}}  \sim 13.5,
\end{eqnarray}
being this value in the range reported by phenomenological investigations~\cite{GPS}.
As discussed above, magnetic field
corrections to $G_{ps}$ have been estimated to be ${\cal O}
\left(\frac{eB}{M^2}\right)^2$ 
\cite{Braghin2020a} and are, therefore, subleading in this regime.
For the mixing angle
we will consider the following value in the range of values collected by PDG
\cite{PDG}:
\begin{eqnarray}
  \theta_{ps}  =  - 20^\circ  .
\end{eqnarray}
To the extent of our knowledge, the only estimation for the dependence of the mixing
angle on the magnetic field has been reported in~\cite{NJL-2022}, and it is very small, so it has not been considered
in this work.

All the figures have been
 produced for fixed total energies $s = 0.036, 0.36, 1.09$ GeV$^2$ and
$20.05$ GeV$^{2}$ for the case of the pion, and
$s= 1.09$ and
$20.05$ GeV$^{2}$ for the $\eta$ and $\eta'$, i.e. only on shell resulting mesons have been considered.

Figures will refer to   probability densities  of gluon fusion into a meson, $P_i$,
in the vacuum ($vac$) and in a finite magnetic field $B$, and their ratios as functions 
of the transverse momentum $p_T$ as defined in Eqs.~\eqref{pT-a} and~\eqref{pT-b}, respectively, for kinematical cases $\alpha = a,b$, respectively. 
They will be denoted as
\begin{eqnarray}
\label{labels-figs-y}
P^{vac} = | T^{gg \to P_i, vac}_{\alpha} |^2,
\;\;\;\;\;
P^B = | T^{gg \to P_i, B}_{\alpha} |^2,
\;\;\;\;\;
\frac{P^B}{P^{vac}} = 
\frac{ | T^{gg \to P_i, B}_{\alpha} |^2}{ | T^{gg \to P_i, vac}_{\alpha} |^2},
\end{eqnarray}
for $P_i =\pi^0, \eta$ and $\eta'$ for $i=0,3,8$, respectively.
The gluon fusion differential rate calculated in Eq.~\eqref{eq:diffrate} will be 
denoted by $R^{vac}_{\eta, \eta'}$.
The kinematical variables, in addition to the angles involved in the process, 
also define the range of the rapidity 
as seen in Eqs.~\eqref{y-a} and \eqref{y-b} for cases (a) and (b), respectively.

 Finally, we will adopt a value 
for the weak magnetic field in the 
limit of validity of this approximation, i.e.
$$
\frac{|eB|}{M^2} = 0.1.
$$
For $M=300$ MeV, this limit corresponds to
$|eB| \sim m_\pi^2$.
As described above, all 
the amplitudes are proportional to $|eB|/M^2$, so that it is straightforward to foresee the behavior for lower magnetic fields with some information about the effect of a slightly different quark effective mass due to a different value of the magnetic field.

\subsection{ Ratios with averaged angles for the pion, 
$y = 0.1$ and $y=0.3$ }
\begin{figure}[H]
\centering
\begin{minipage}{.4\textwidth}
  \centering \includegraphics[width=.95\linewidth]{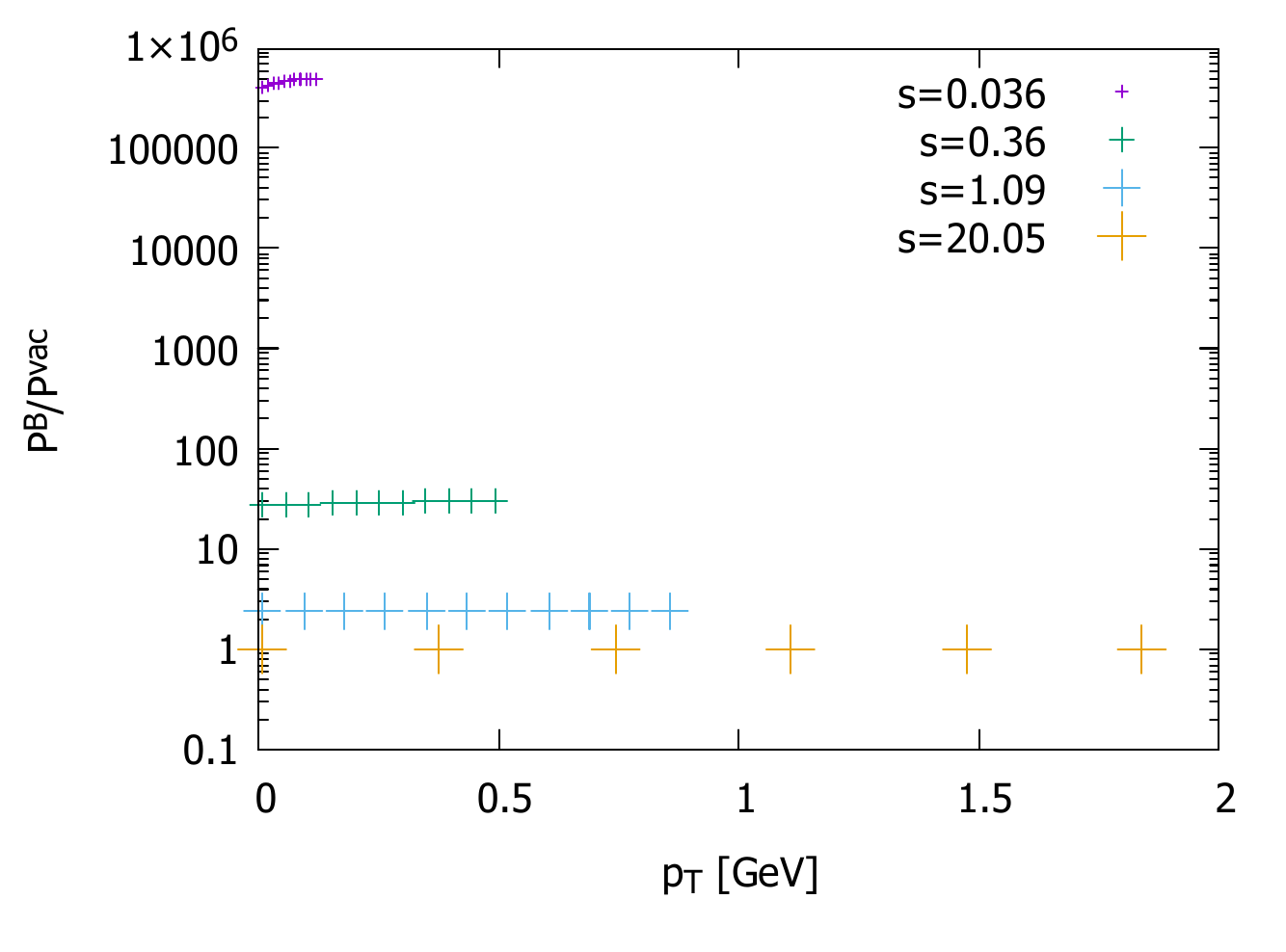}
  \caption{  $\frac{|T^{gg \to \pi^0, B}_a |^2}{
|T^{gg \to \pi^0, vac}_a |^2
}$.
Case a, quark masses of  Set I in \eqref{M300},   averaged angles for $y=0.1$.}
  \label{fig:ratio-pi0-y01}
\end{minipage}%
\hspace{1cm}
\begin{minipage}{.4\textwidth}
  \centering

\includegraphics[width=.95\linewidth]{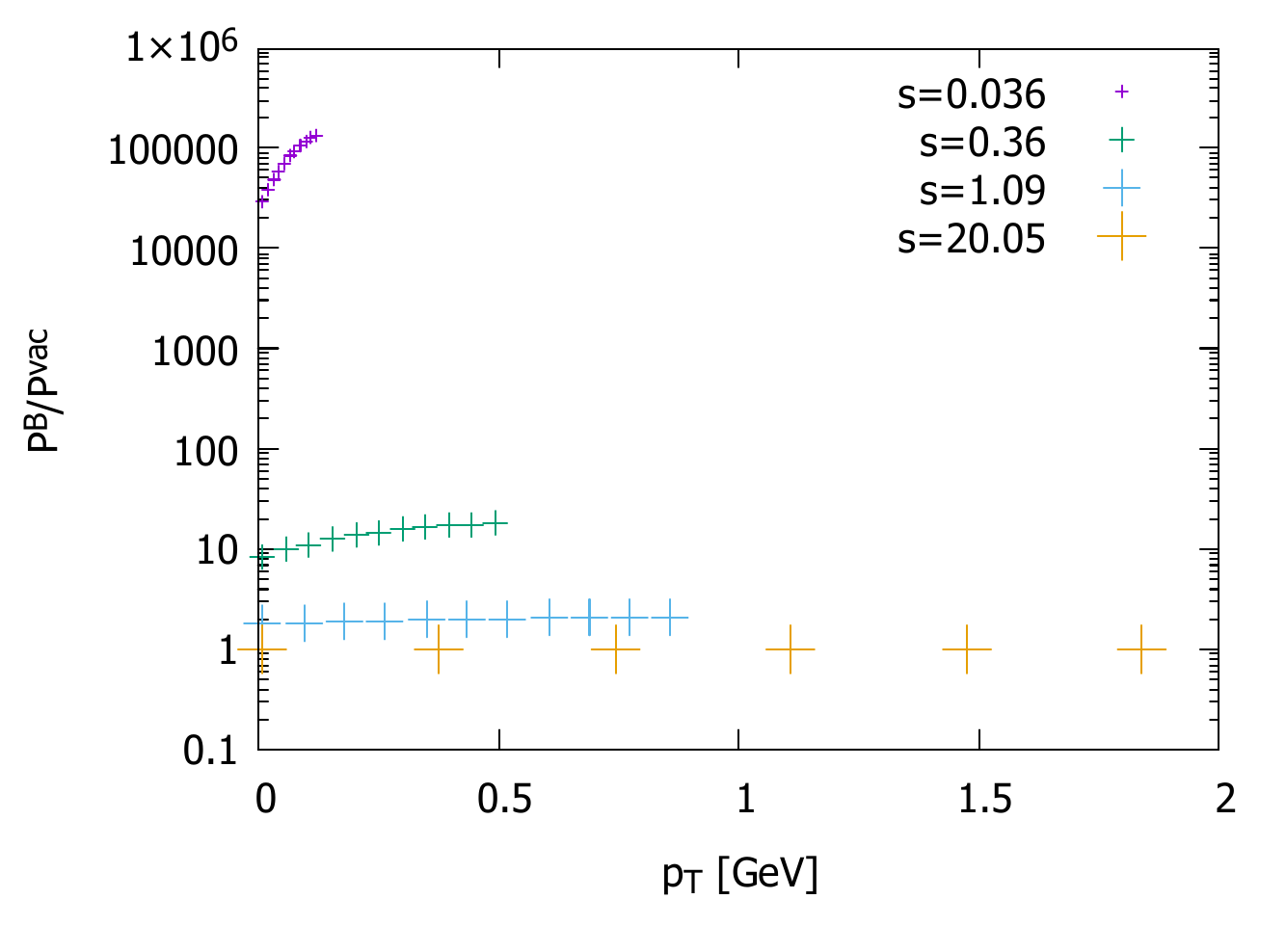}
  \caption{$ \frac{|T^{gg \to \pi^0, B}_a |^2}{
|T^{gg \to \pi^0, vac}_a |^2
}$.
Case a, quark masses of  Set I in \eqref{M300},   averaged angles for $y=0.3$.}
  \label{fig:ratio-pi0-y03}
\end{minipage}
\end{figure}

Firstly, we report results for the case of gluon fusion into a pion with averaged angles $\phi$, $\varphi$, and set I of quark effective masses.
In Figs.~\eqref{fig:ratio-pi0-y01} 
and \eqref{fig:ratio-pi0-y03}
the ratio of probability densities 
$P^B/P^{vac}$, defined in Eq.~\eqref{labels-figs-y},
is exhibited for 
case (a) - Section~\eqref{sec:case-a} - and different total incoming gluon energies, for 
$y=0.1$ and $y=0.3$, respectively.
It is seen that, for very low energies,
 i.e. just above the threshold energy 
for pion production, 
there is an extremely large increase in the 
 rate of gluon fusion into pion due to the relatively weak magnetic field,
as compared to the rate for the process in the vacuum.
Nevertheless, this 
ratio  
decays very strongly with energy, so
that at $s \sim 1\,\text{GeV}^2$, the ratio drops to just above $\sim 1$, and the magnetic enhancement effect is then lost.
The overall dependence with energy goes 
with $1/s^n$ for $n=2,3,4$ for the different integrals in the magnetic field dependent part, which explains the strong enhancement at low energies $s \lesssim 1~\text{GeV}^2$, in contrast with the suppression of this effect for $s \gtrsim 1~\text{GeV}^2$.
It is also interesting to note that 
the gluon fusion ratio at  
$y=0.1$ is larger 
than   at  $y=0.3$.
At small energies, there is a larger variation of amplitudes  
(in the presence of magnetic field) with $p_T$, 
and this can be seen in 
Eqs.~\eqref{pT-a}.
As the gluon energies increase, the dependence of the ratio on $p_T$ progressively disappears,
until they tend to reach an asymptotic limit independent on $p_T$ at larger energies.
This is more pronounced for 
$y=0.3$, whereas for 
$y=0.1$ a milder dependence on $p_T$ is observed, even in the low $p_T \lesssim 0.3$~GeV range.
 For the analysis of the  angular distribution of emitted pions (and 
  $\eta$ and $\eta'$) we would need,
  according for example to Eqs.~\eqref{pT-a}
  and \eqref{pT-b},
  some knowledge of the gluons energies for a fixed direction of the magnetic field, which will not be discussed in this work.

\begin{figure}[H]
\centering
\begin{minipage}{.4\textwidth}
  \centering \includegraphics[width=.9\linewidth]{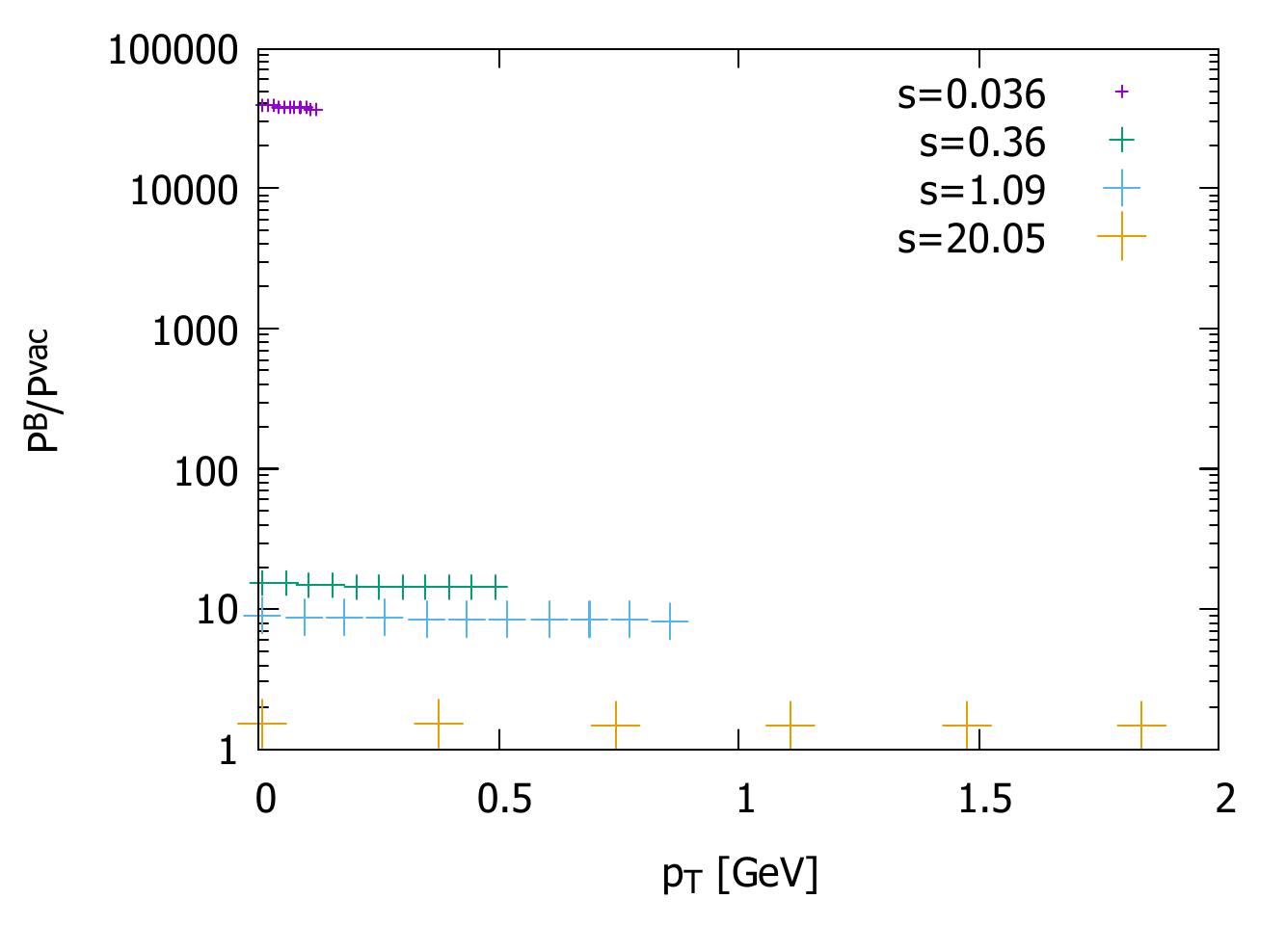}
  \caption{  $\frac{|T^{gg \to \pi^0, B}_b |^2}{
|T^{gg \to \pi^0, vac}_b |^2
}$.
Case b, quark masses of  Set I in \eqref{M300},   averaged angles for $y=0.1$.} 
  \label{fig:ratio-pi0-b-y01}
\end{minipage}%
\hspace{1cm}
\begin{minipage}{.4\textwidth}
  \centering
  \includegraphics[width=.9\linewidth]{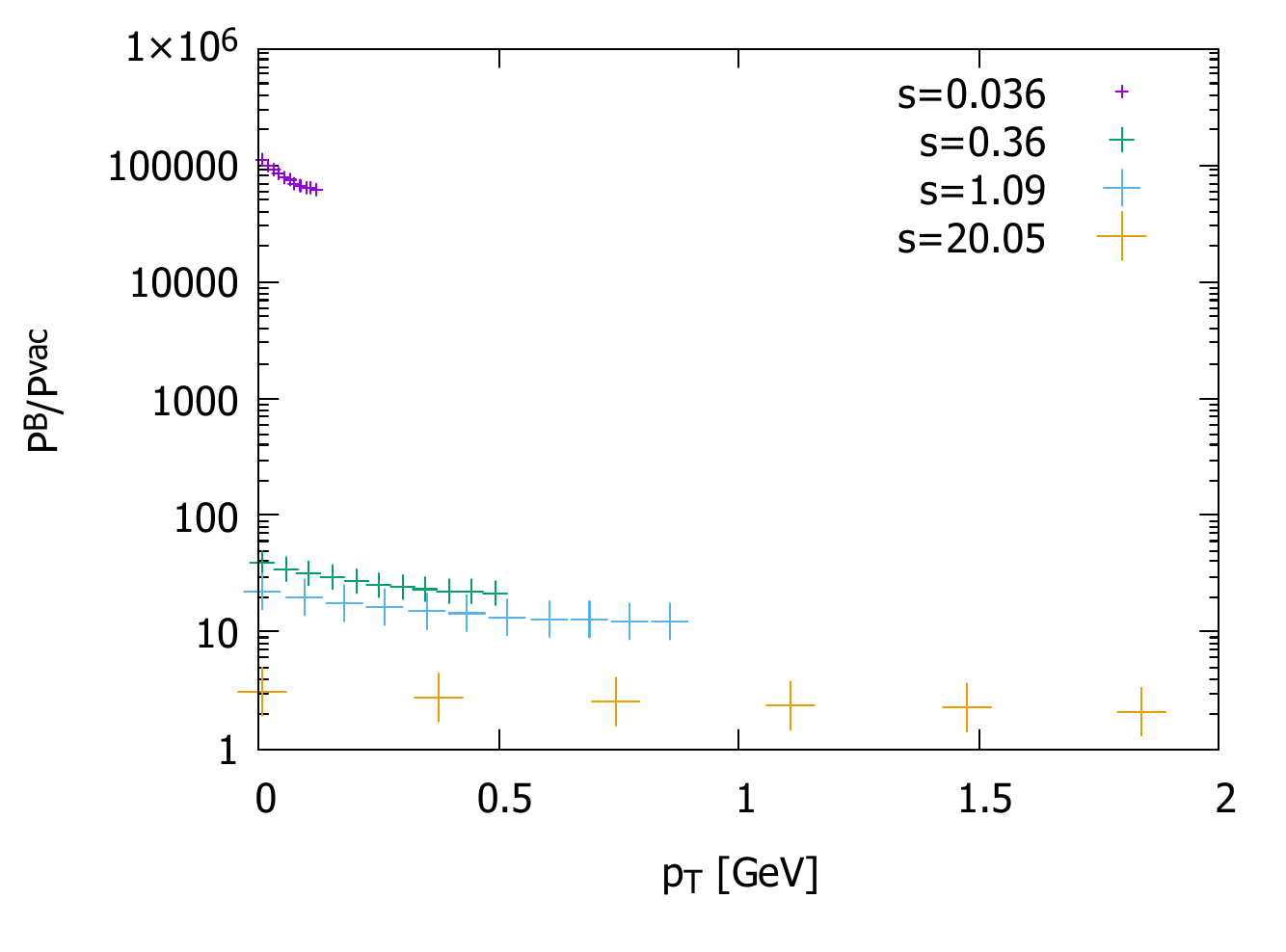}
  \caption{$ \frac{|T^{gg \to \pi^0, B}_b |^2}{
|T^{gg \to \pi^0, vac}_b |^2
}$.
Case b, quark masses of  Set I in \eqref{M300},   averaged angles for $y=0.3$.}
  \label{fig:ratio-pi0-b-y03}
\end{minipage}
\end{figure}

In 
 Figs. \eqref{fig:ratio-pi0-b-y01} 
and \eqref{fig:ratio-pi0-b-y03}, we represent the same quantities already discussed for case (a), but now for the kinematics of case (b).
The overall  behavior with gluon energies  is similar, 
although the ratios tend to be smaller (larger)
 for case (b) for small (larger)  energies.
Besides that, the ratios decrease slightly for very small $p_T$ instead of increasing as it occurs in case (a).
The asymptotically flat behavior for higher energies is reached faster for 
$y=0.1$ than for $y=0.3$.

\begin{figure}[H]
\centering
\hspace{1cm}
\begin{minipage}{.4\textwidth}
  \centering
  \includegraphics[width=.9\linewidth]{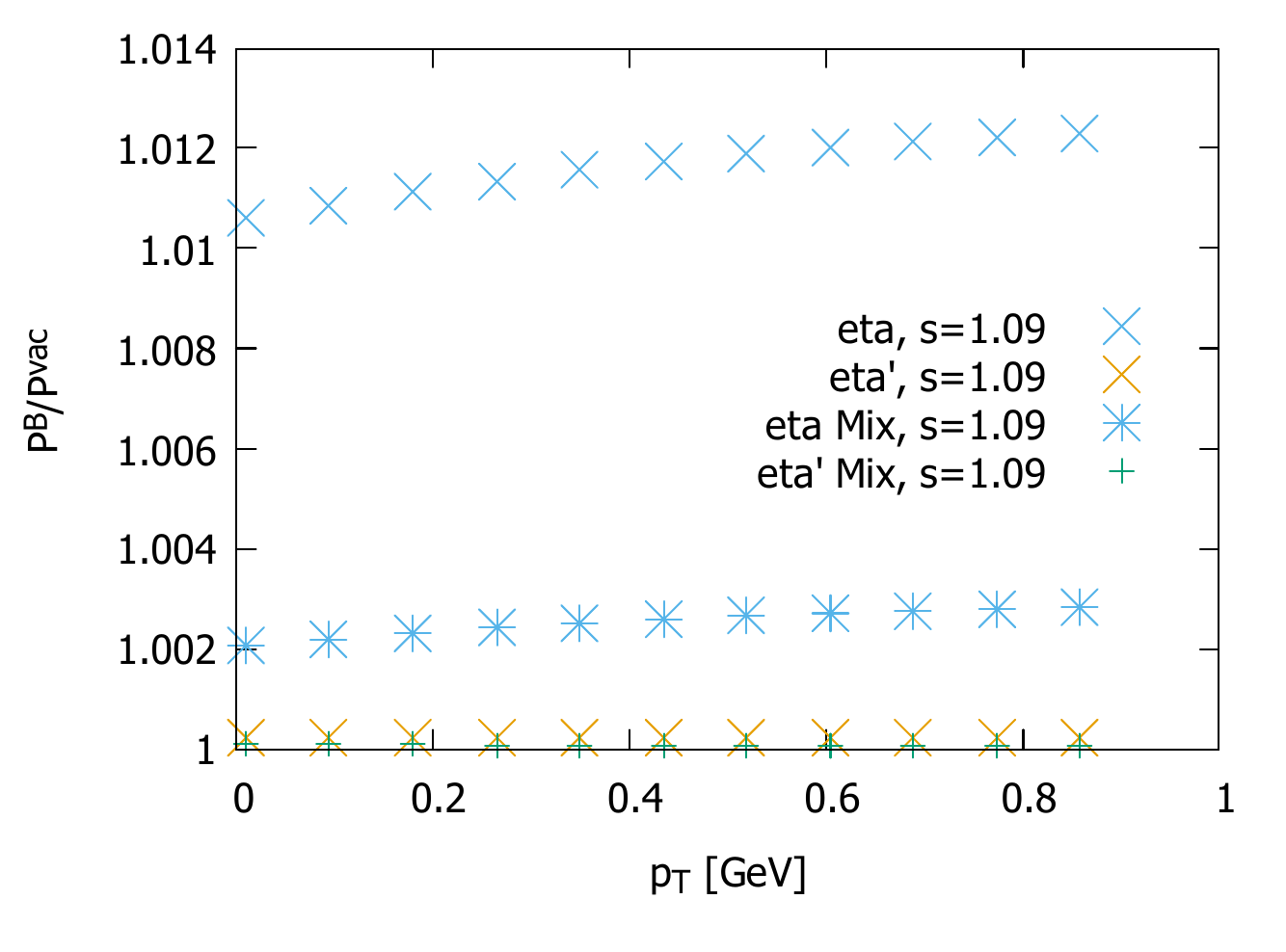}
  \caption{$ \frac{|T^{gg \to \eta,\eta', B}_a |^2}{
|T^{gg \to  \eta,\eta', vac}_a |^2
}$.
Case a, quark masses of  Set I in \eqref{M300},   averaged angles for $y=0.3$.}
  \label{fig:ratio-etas-y03}
\end{minipage}
\begin{minipage}{.4\textwidth}
  \centering
  \includegraphics[width=.9\linewidth]{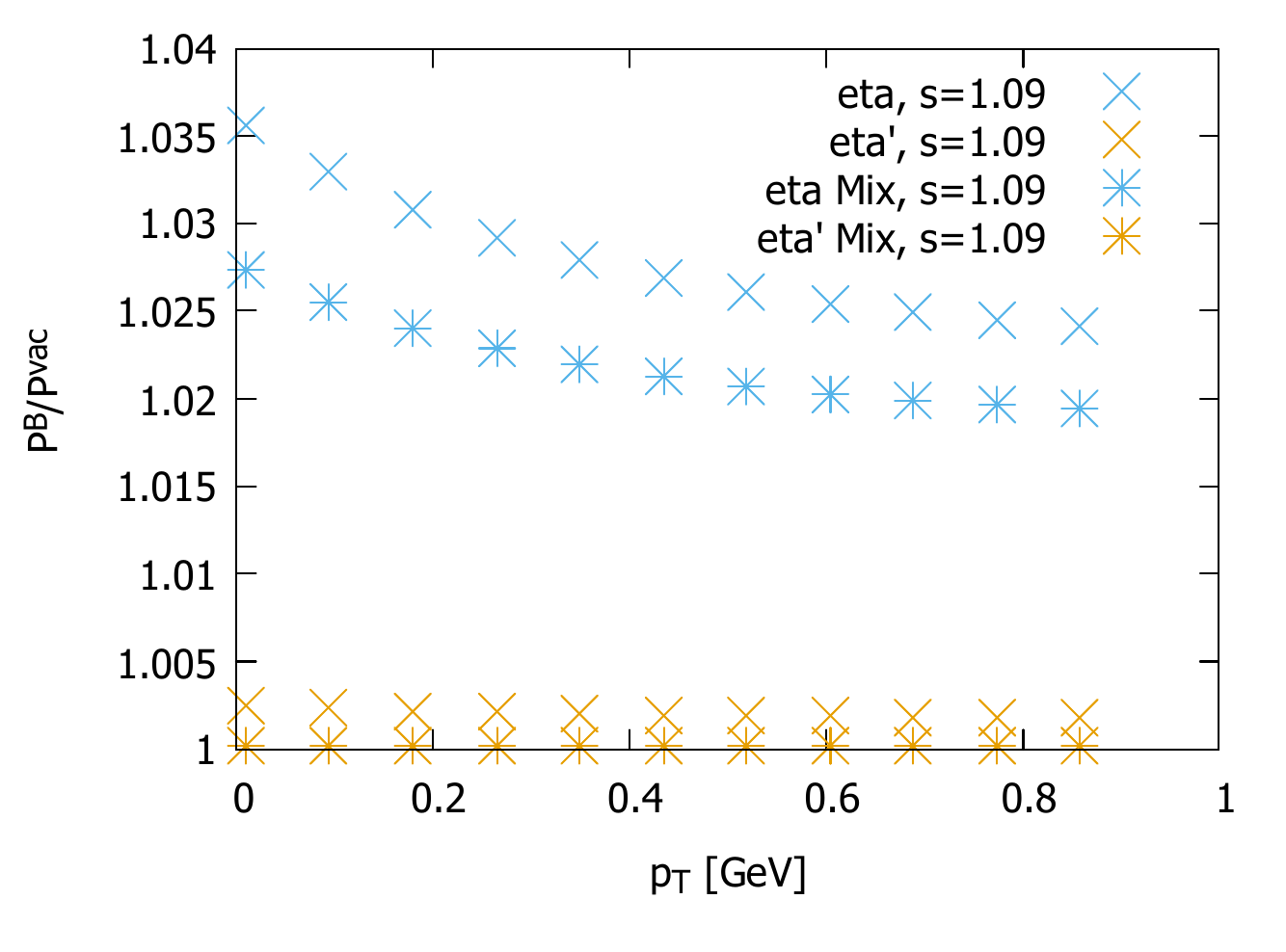}
  \caption{$ \frac{|T^{gg \to \eta,\eta', B}_b |^2}{
|T^{gg \to  \eta,\eta', vac}_b |^2
}$.
Case b, quark masses of  Set I in \eqref{M300},   averaged angles for $y=0.3$.}
  \label{fig:ratio-etas--b-y03}
\end{minipage}
\end{figure}

As it is seen in the last figures, 
the cases in which $y=0.3$ present less
flat ratios.
Therefore, to facilitate visualization, in Figs.~\eqref{fig:ratio-etas-y03} and \eqref{fig:ratio-etas--b-y03}
we choose $y=0.3$ to present the ratios 
of gluon fusion probability density into $\eta$ and 
$\eta'$, under magnetic field, over its values
in the vacuum, 
as functions of $p_T$
at the energy of $s=1.09$ GeV$^2$, 
i.e. above the threshold of production of $\eta'$,
respectively for case (a) and (b).
The same ratios are also presented, in these figures, 
by considering the meson-mixing
of Eq. \eqref{meson-mixing}.
There is a smooth increase (or decrease) of the ratio 
for increasing $p_T$
for case (a) ( or (b)).
 Whereas the meson mixing brings strength 
from the $\eta'$ to the $\eta$
in the vacuum, at finite 
magnetic field the opposite effect occurs.
The relative changes are different in both kinematic cases. Case (a) exhibits a very mild relative increase in the $\eta$ production due to the magnetic field, that grows with $p_T$, with the $\eta'$ production almost unaffected. For case (b), in contrast, while a relative enhancement in the production probability for $\eta$ due to the magnetic field is still observed, this effect decreases with $p_T$, with an almost negligible effect on $\eta'$. The same trend is also observed when mixing is incorporated, but with a comparatively even milder magnetic enhancement in both cases.

\begin{figure}[H]
\centering
\begin{minipage}{.4\textwidth}
  \centering \includegraphics[width=.9\linewidth]{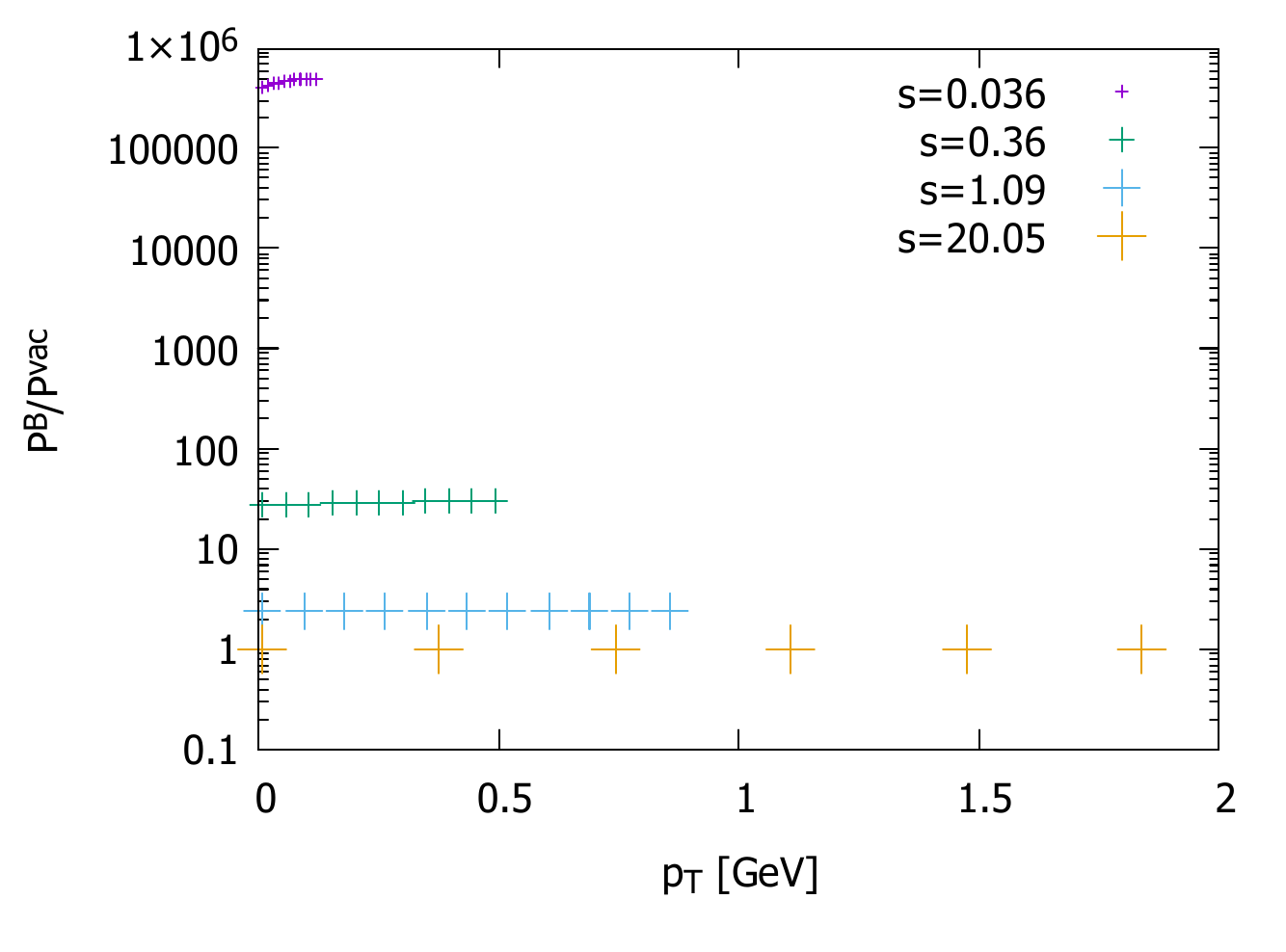}
  \caption{  $\frac{|T^{gg \to \pi^0, B}_a |^2}{
|T^{gg \to \pi^0, vac}_a |^2
}$.
Case a, quark masses of  Set II in \eqref{M330},   averaged angles for $y=0.1$.}
  \label{fig:ratio-pi0-y01M330}
\end{minipage}%
\hspace{1cm}
\begin{minipage}{.4\textwidth}
  \centering
  \includegraphics[width=.9\linewidth]{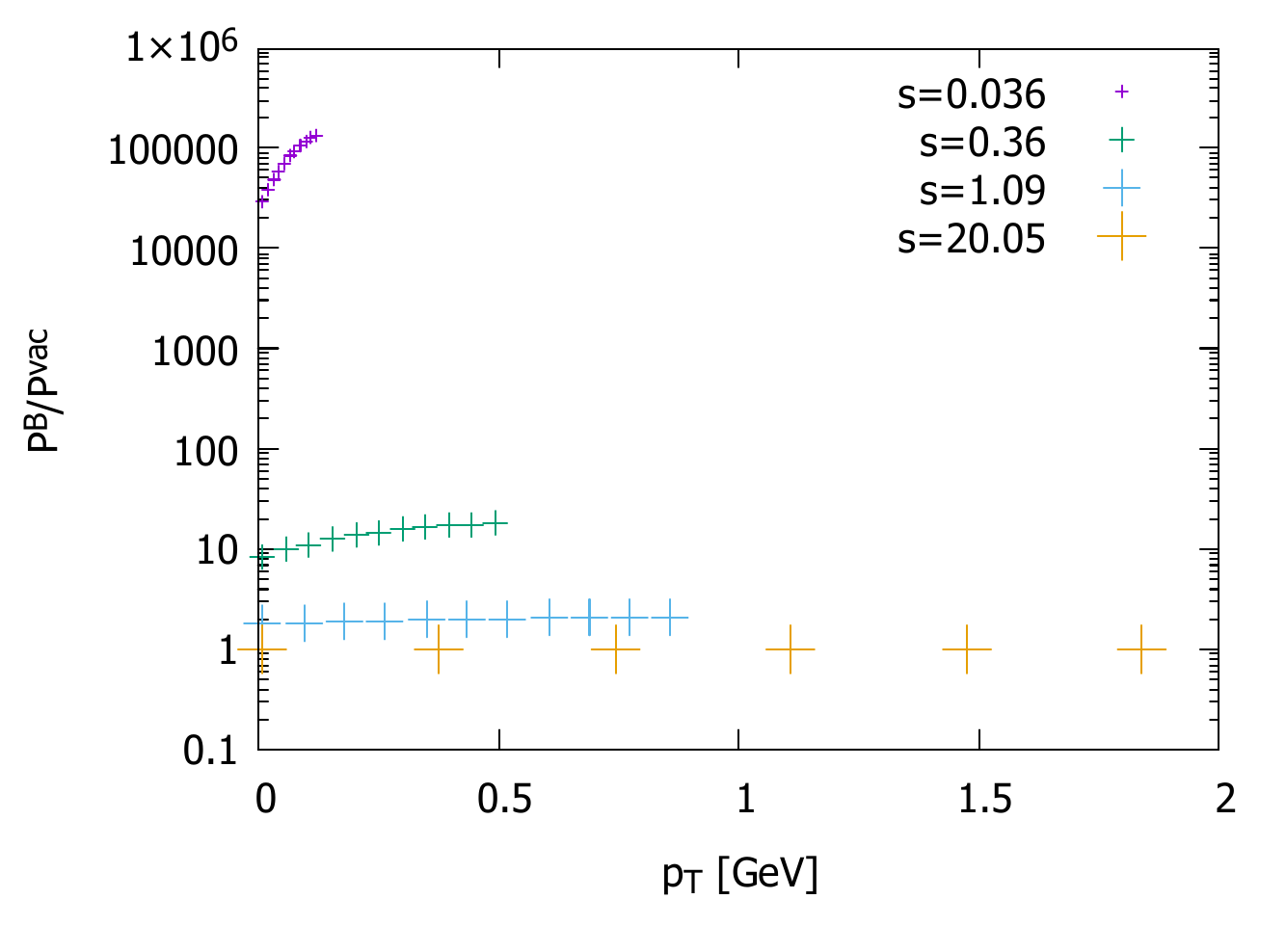}
  \caption{$ \frac{|T^{gg \to \pi^0, B}_a |^2}{
|T^{gg \to \pi^0, vac}_a |^2
}$  .
Case a, quark masses of  Set II in \eqref{M330},   averaged angles for $y=0.3$.}
  \label{fig:ratio-pi0-y0M3303}
\end{minipage}
\end{figure}

Since the constituent quark masses are strongly model dependent, in 
Figs. 
\eqref{fig:ratio-pi0-y01M330} and \eqref{fig:ratio-pi0-y0M3303}
the ratios addressed above,
$P^B/P^{vac}$, for the neutral pion is presented 
for the same energies 
previously considered, as functions of $p_T$,
but for the set II of constituent quark masses \eqref{M330}, respectively for case (a) for the two rapidities considered above.
The ratios are slightly larger than those from
Figs. 
\eqref{fig:ratio-pi0-y01} 
and \eqref{fig:ratio-pi0-y03}
for the set I of constituent quark masses.
The ratios go roughly with $\sim (330/300)^n$ for $n=1,2$.
The variation of the ratio with $p_T$ is much stronger for $y=0.3$ than
for $y=0.1$.

\subsection{ Gluon fusion  rates for $\eta$ and $\eta'$   in the vacuum}

\begin{figure}[H]
\centering
\begin{minipage}{.4\textwidth}
  \centering \includegraphics[width=.9\linewidth]{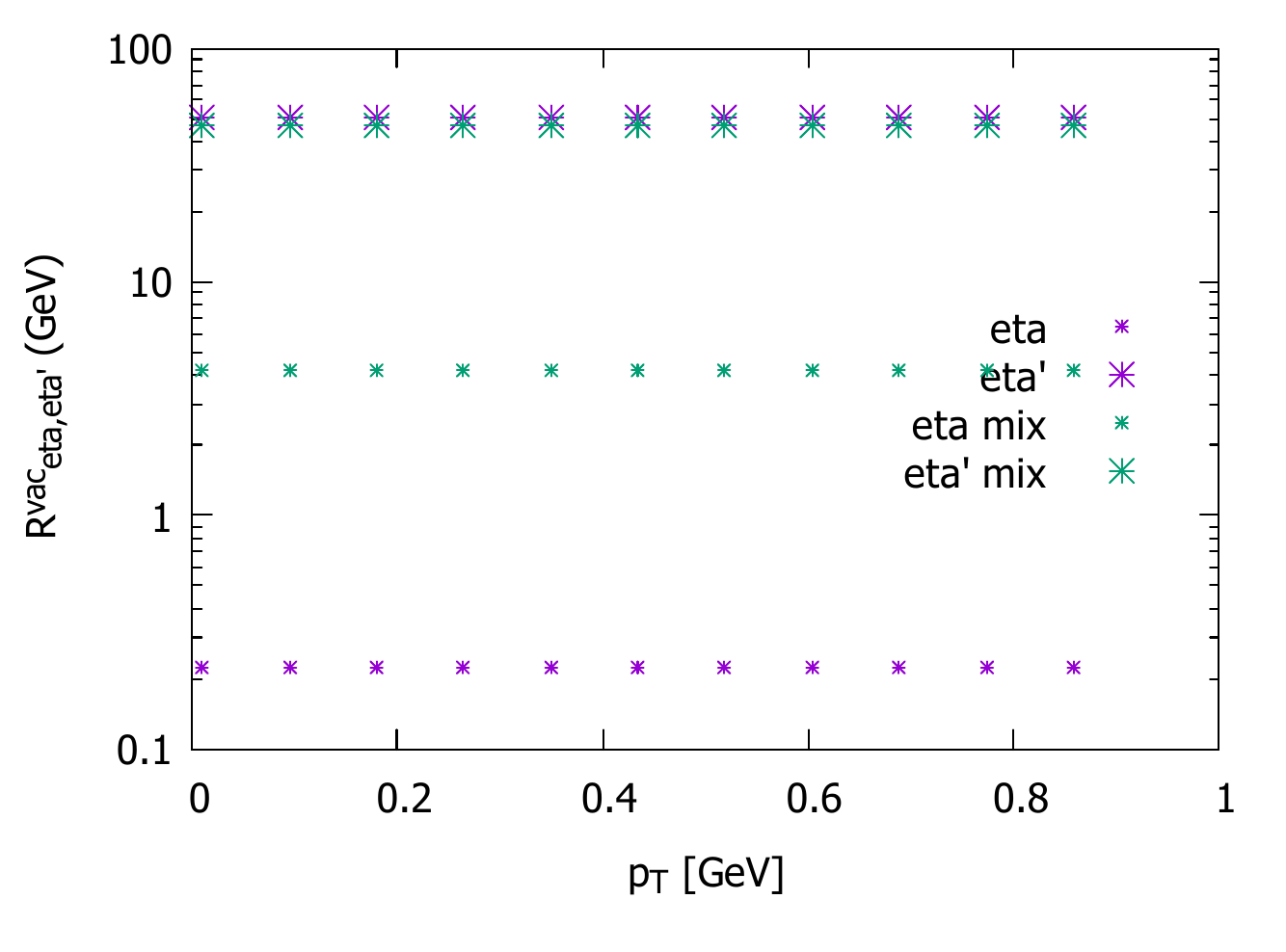}
  \caption{ 
 Gluon fusion rate
  into $\eta, \eta'$ in the vacuum
  for
case a and averaged angles, for $y=0.1$
and set I in \eqref{M300}
at $s=1.09$ GeV$^2$.}
  \label{fig:rate-eta109-a-y01}
\end{minipage}%
\hspace{1cm}
\begin{minipage}{.4\textwidth}
  \centering
\includegraphics[width=.9\linewidth]{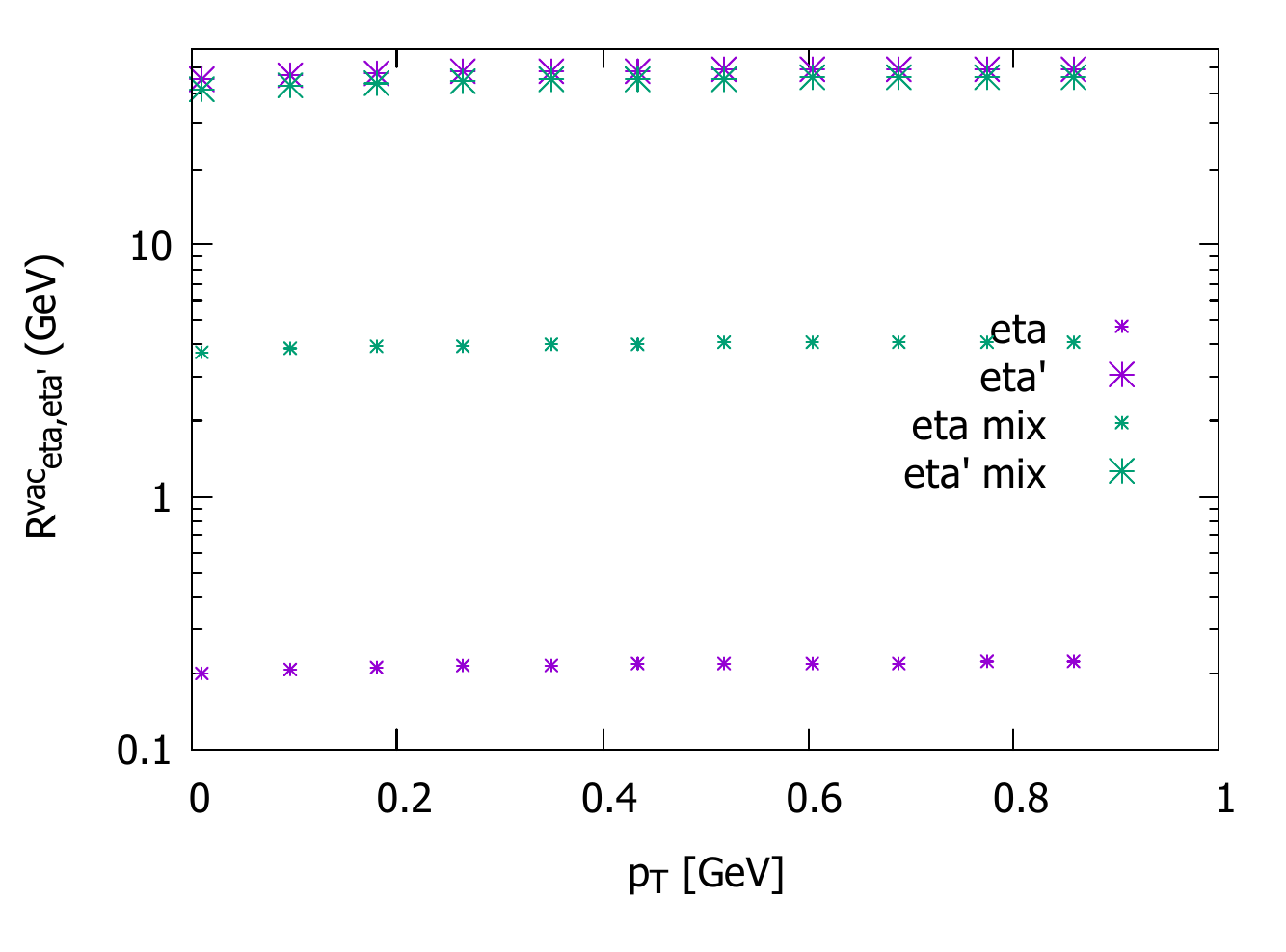}
  \caption{ Gluon fusion rate
  into $\eta, \eta'$ in the vacuum
  for
case a and averaged angles, for $y=0.3$
and set I in \eqref{M300}
at $s=1.09$ GeV$^2$.}
  \label{fig:rate-eta109-a-y03}
\end{minipage}
\end{figure}

\begin{figure}[H]
\centering
\begin{minipage}{.4\textwidth}
  \centering \includegraphics[width=.9\linewidth]{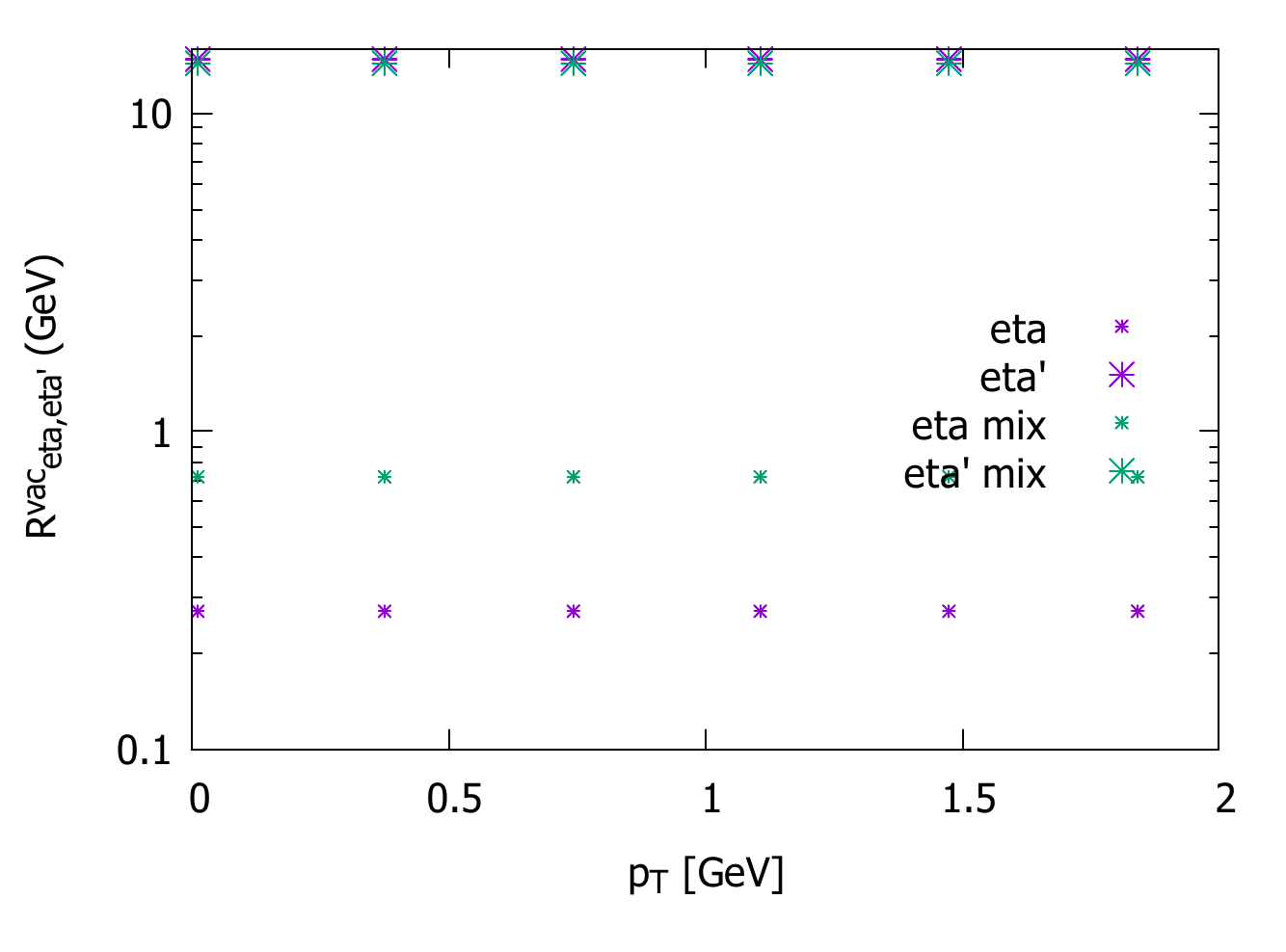}
  \caption{
   Gluon fusion rate
  into $\eta, \eta'$ in the vacuum
  for
case a and averaged angles, for $y=0.1$
and set I in  \eqref{M300}
at $s=20.05$ GeV$^2$.}
  \label{fig:rate-eta2005-a-y01}
\end{minipage}%
\hspace{1cm}
\begin{minipage}{.4\textwidth}
  \centering
  \includegraphics[width=.9\linewidth]{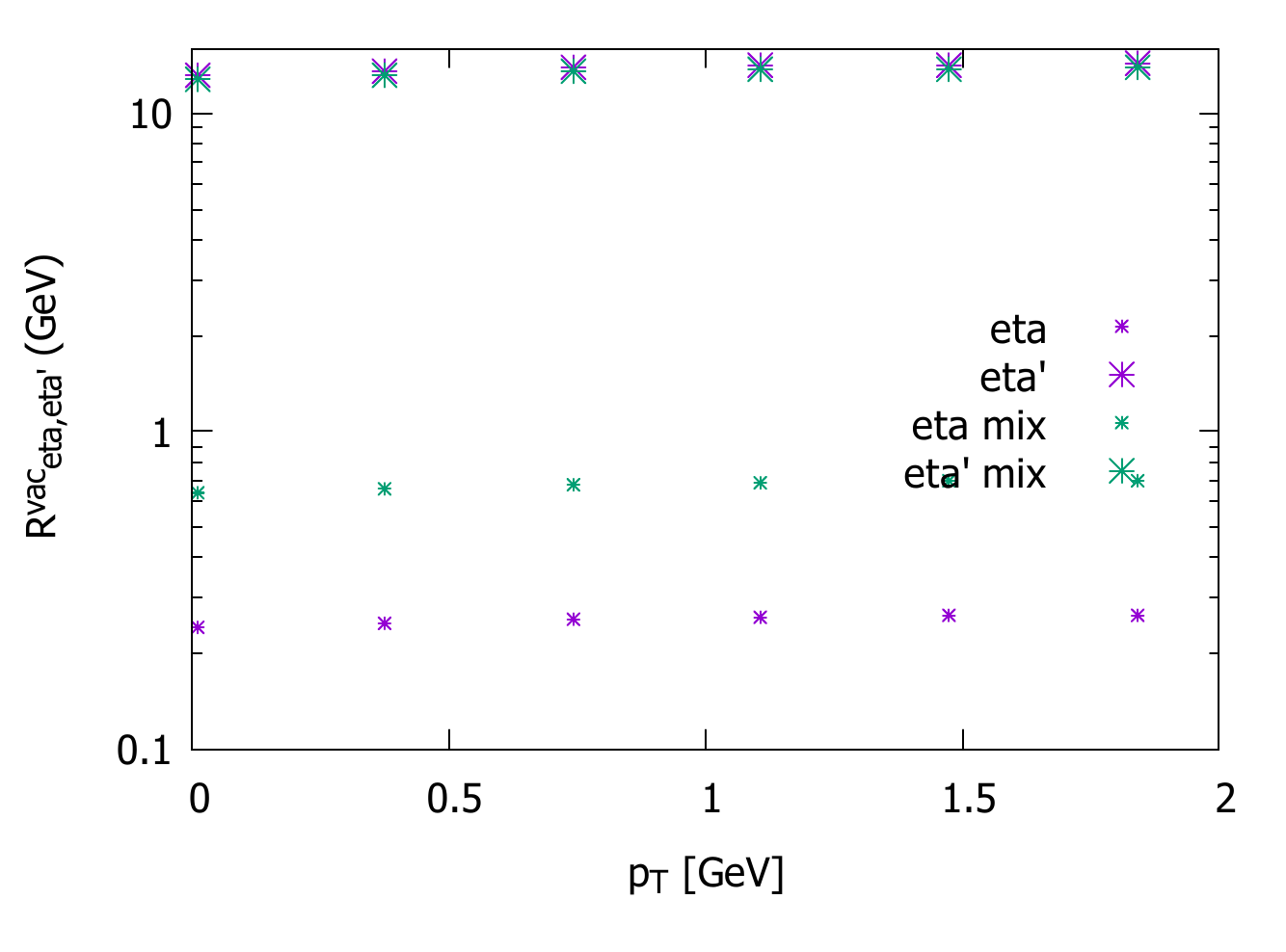}
  \caption{ Gluon fusion rate
  into $\eta, \eta'$ in the vacuum
  for
case a and averaged angles, for $y=0.3$
and set I in \eqref{M300}
at $s=20.05$ GeV$^2$.}
  \label{fig:rate-eta2005-a-y03}
\end{minipage}
\end{figure}

The ratio of the magnetic field dependent rate of $\eta$ and $\eta'$ with 
respect to the vacuum part is small, so we plot next the normalized
 rates of gluon fusion in the 
vacuum as defined in Eqs.~\eqref{diff-rate}.
In Figs.~\eqref{fig:rate-eta109-a-y01} and \eqref{fig:rate-eta109-a-y03},
 these rates  
in the vacuum are  shown for case (a) at 
$s=1.09$ GeV  with $y=0.1$ and $y=0.3$, respectively, for 
set  I of constituent quark masses.
Their behavior is nearly independent of $p_T$, as seen by the flat curves, consistent with the previous results discussed for the 
pion, as shown in the previous section.
The $\eta-\eta'$ mixing changes slightly the total fusion rates,
being the log-scale responsible for unequal shifts for the $\eta$
and $\eta'$ cases.
In Figs.~\eqref{fig:rate-eta2005-a-y01} and \eqref{fig:rate-eta2005-a-y03}, we show
the same rates of previous two figures, but for
$s=20.05$ GeV$^{2}$.
The behavior is the same as for the 
case presented in the previous figures at $s=1.09$~GeV$^2$, but the value is nearly one order of magnitude smaller.

The behavior of fusion rates into $\eta$ and $\eta'$ are the same for 
case (b), being the absolute values smaller, so they will not be shown.
The pattern for slightly larger 
constituent quark masses, such as for 
case II, is the same with a small increase in the absolute values of decay rates, 
and then the corresponding figures are not 
exhibited here.

%


\section{Discussion and Conclusions}
\label{conclusion}
In this work, we investigated the gluon fusion process as a possible mechanism for neutral meson generation, concretely $\pi^0$, $\eta$, and $\eta'$, in a non-thermalized QGP. The large amount of gluons generated might lead to the possible scenario of the color glass condensate~\cite{McLerran2002,mclerran2008colorglasscondensateglasma}. Our model represents a phenomenological approximation to the conditions generated in the early stages of HIC, prior to thermalization, when a background magnetic field is presumably generated, particularly in non-central collisions. In this context, we compared the relative probabilities for neutral meson emission when the magnetic field is present, with respect to the vacuum (zero-field) case, as a function of transverse momenta $p_T$ and energy $\sqrt{s}$, close to midrapidity. The ratio of the rates for these cases, corresponding to 
$\frac{|T^{gg\to P_i,B}|^2
}{
|T^{gg\to P_i,vac}|^2}$, (for $P_i = \pi^0$, $\eta$, or $\eta'$) has components roughly proportional to the inverse 
of total energy to some power, i.e.
$1/s^n$, with additional logarithmic dependence.

Interestingly, our results reveal that for neutral pion production, the magnetic field produces a very strong catalytic effect, which is more dramatic in the  low $s < 1~\text{GeV}^2$ range. In contrast, the relative enhancement of the $\pi^0$ emission due to the finite magnetic field tends to disappear. We remark that here is a small window for $p_T \lesssim  0.3~\text{GeV}$ where the pion emission probability ratio exhibits a strong, monotonically decreasing (or increasing) dependence on the transverse momentum. Beyond this region, the emission probability ratio tends to an asymptotically momentum-independent value. We clearly identify a different dependence of the emission rate at small values of $p_T$, for the two different configurations Case~(a) and Case~(b) considered in this work, and the reason can be traced back to kinematics. For Case~(a), Eq.~\eqref{eq_tanycasea} indicates that, for $p_T \ll \sqrt{s}$, the scattering angle satisfies $\tan\theta \propto p_T^{-1}$, and hence the emission probability grows as $\cos\theta \propto p_T$ for very low transverse momenta. In contrast, for Case (b), Eqs.~\eqref{eq_ECaseb}-\eqref{eq_tanycaseb} imply that for $p_T \ll \sqrt{s}$, $\tan\theta \propto p_T$, thus implying that the emission probability grows as $\cos\theta\ \propto p_T$ for very low transverse momenta.
The kinematical regimes of cases (a) and (b) lead to different behaviors
as concerning the dependencies on the (small) rapidities, 
i.e. the probability ratio might slightly increase (case (b)) or decrease  (case(a))
when going from  $y=0.1$ to $y=0.3$.
This large amplification of very low energy pion emission is consistent with 
the scenario of (instantaneous) neutral pion condensation.

Given that the threshold for the 
gluon fusion into a $\eta'$ is nearly $s\sim1$~GeV$^2$, the 
probability ratios for the $\eta$ and $\eta'$ are already lowered by
the factors $1/s^n$. Besides that, 
their quark-structure when coupled to the magnetic field, 
as seen in Eqs.~\eqref{mesons-Bfield}, already leads to a suppression due to the small isospin and flavor symmetry breaking.
On the other hand, the dependence on the transverse momentum is very weak for this case, with kinematics controlling the  magnitude: 
The emission probability ratios for Case~(b) are  approximately one order of magnitude larger than those for Case~(a). In the case of $\eta$ and $\eta'$ mesons, we also analyzed its mixing, as parameterized by the mixing angle $\theta_m$. For this second case, we observe that the emission probability ratio exhibits a similar trend, by with an overall smaller magnitude. As we mentioned in the Introduction, experimental data~\cite{Acharya_2018} exhibits some tension with the hydrodynamic and pQCD hadronization models for $3 < p_T < 6$~GeV. Even though no experimental data is available yet in the very small range $p_T < 1$~GeV, our theoretical results suggest that the interplay between magnetic field and kinematics in this very small $p_T$ region may reflect the presence of the proposed gluon fusion mechanism.

In this work, we studied the gluon fusion process for two different, and mutually independent, kinematic configurations, that were assumed {\it{a priori}}. However, in order to directly connect our theoretical model with experimental data, one could combine it with the gluon energy or momentum distributions, thus leading to more precise predictions of angular distributions for the emitted neutral pseudoscalar mesons in the presence of a background magnetic field. This information may lead to a determination of the magnetic fields in ultrarelativistic heavy-ion collisions (mainly non-central cases), which have been shown to be quite elusive so far due to the inherent complexity of the physical system and its kinematics/dynamics. Furthermore, one may foresee that the angular distributions of detected pions could be helpful to identify those emitted from this proposed gluon fusion mechanism involving the presence of the magnetic field. These are future directions that we are currently developing. In addition, as an alternative for possible signal detection, we will analyze possible consequences 
for the 
(direct) photon emission of the emitted PS mesons, to be communicated in a forthcoming work.

\section*{Acknowledgments}

F.L.B.  
thanks the hospitality of 
E.M. and the Faculty of Physics of PUC - Chile  during his visit.
F.L.B.
acknowledges partial financial support from 
CNPq-307792/2025-0, and ANID Fondecyt Grant No 1230440 for his visit.
E.M. acknowledges financial support from ANID Fondecyt Grant No 1230440.

\bibliography{References}

\appendix

\section{ kinematical variables}
\label{App:kinematics}

In this Appendix, we present the details to obtain the kinematic relations for cases (a) and (b), respectively. We also present the derivation of the expressions that allow to convert the set of independent variables $\left( E_1, E_2, \theta \right)$, into $\left(s,p_{T},y  \right)$. 

\subsection{Case (a)}
Let us first consider {\bf{case (a)}}, with
\begin{eqnarray}
P_1^{\mu} &=& (E_1,0,0,-|\mathbf{P}_1|)\nonumber\\
P_2^{\mu} &=& (E_2, |\mathbf{P}_2|\sin\theta\cos\varphi,|\mathbf{P}_2|\sin\theta\sin\varphi,|\mathbf{P}_2|\cos\theta)
\end{eqnarray}
For on-shell gluons $P_1^2 = P_2^2 = 0$, we have
\begin{eqnarray}
\left|\mathbf{P}_1\right| = E_1,\,\,\,\, \left|\mathbf{P}_2\right| = E_2,
\end{eqnarray}
and then
\begin{eqnarray} 
P_1\cdot P_2 = E_1 E_2 \left( 1 + \cos\theta \right) \equiv \frac{s}{2}
\end{eqnarray}
Then, we define the transverse and longitudinal momenta as follows
\begin{eqnarray}
\label{pT-a}
p_T &=& P_{1,z} + P_{2,z} =  E_2\cos\theta - E_1 \nonumber\\
p_{\parallel} &=& \sqrt{\left(P_{1,x} + P_{2,x}\right)^2 + \left(P_{1,y}+P_{2,y}\right)^2} = E_2\sin\theta
\end{eqnarray}
We also introduce the definition of the rapidity $y$, such that
\begin{eqnarray}
\label{y-a}
y = \frac{1}{2}\ln\left( \frac{P^0 + p_{\parallel}}{P^0 - p_{\parallel}} \right)\Longrightarrow \frac{p_{\parallel}}{P^0} = \frac{e^{2 y}-1}{e^{2 y}+1} = \tanh(y).
\end{eqnarray}
Summarizing, the above definitions lead to the system of equations that determine the angle $\theta$, and the incoming gluon energies $E_1,E_2$ in terms of $y$, $s$, and $p_T$, as follows
\begin{equation}
\frac{E_2\sin\theta}{E_1 + E_2} = \tanh(y)
\label{eq_A}
\end{equation}
\begin{equation}
E_2\cos\theta = E_1 + p_{T}
\label{eq_B}
\end{equation}
\begin{equation}
E_2 (1 + \cos\theta) = \frac{s}{2 E_1}
\label{eq_C}
\end{equation}

The kinematic variables $\left(E_1,E_2,\theta\right)$ can be fully expressed in terms of $\left(s,p_T,y \right)$. For this purpose, 
we first combine Eq.~\eqref{eq_A} and Eq.~\eqref{eq_B}, by taking the square 
\begin{eqnarray}
E_2^2\sin^2\theta &=& \left(  E_1 + E_2\right)^2\tanh^2(y)\nonumber\\
E_2^2\cos^2\theta &=& \left(E_1 + p_{T}\right)^2  
\end{eqnarray}
and then adding them, to obtain
\begin{eqnarray}
E_2^2 = \left(  E_1 + E_2\right)^2\tanh^2(y) + \left(E_1 + p_{T}\right)^2.
\label{eq_Dint}
\end{eqnarray}
Then, combining Eq.~\eqref{eq_C} and Eq.~\eqref{eq_B}, we have
\begin{eqnarray}
E_2 = \frac{s}{2 E_1} - E_1 - p_{T},
\label{eq_Dint2}
\end{eqnarray}
and substituting this into Eq.~\eqref{eq_Dint}, we get
\begin{eqnarray}
\left( \frac{s}{2 E_1} - p_{T} \right)^2 + E_1^2 - 2 E_1\left( \frac{s}{2 E_1} - p_{T} \right) = \left( \frac{s}{2 E_1} - p_{T} \right)^2 \tanh^2(y) + \left(E_1 + p_{T}\right)^2.
\end{eqnarray}
After some elementary algebra, this equation yields
\begin{eqnarray}
\frac{s}{2 E_1} - p_{T} = \sqrt{\frac{p_{T}^2 + s}{1 - \tanh^2(y)}}-
\end{eqnarray}
Then, finally solving in favor of $E_1$, we obtain the result
\begin{eqnarray}
E_1 = \frac{s}{2\left( p_{T} + \sqrt{\frac{p_{T}^2 + s}{1 - \tanh^2(y)}} \right)},
\label{eq_E1}
\end{eqnarray}
and from Eq.~\eqref{eq_Dint2}
\begin{eqnarray}
E_2 = \sqrt{\frac{p_{T}^2 + s}{1 - \tanh^2(y)}} - \frac{s}{2\left( p_{T} + \sqrt{\frac{p_{T}^2 + s}{1 - \tanh^2(y)}} \right)}.
\label{eq_E2}
\end{eqnarray}
Finally, the angle is obtained by taking the ratio between Eq.~\eqref{eq_A}, and Eq.~\eqref{eq_B}, as follows
\begin{eqnarray}
\tan\theta = \frac{E_1 + E_2}{E_1 + p_{T}}\tanh(y).
\end{eqnarray}
After substituting $E_1$ and $E_2$ from Eq.~\eqref{eq_E1} and Eq.~\eqref{eq_E2}, we finally obtain
\begin{eqnarray}
\tan\theta = \frac{ \sqrt{\frac{p_{T}^2 + s}{1 - \tanh^2(y)}}  }{ p_{T}+ \frac{s}{2\left( p_{T} + \sqrt{\frac{p_{T}^2 + s}{1 - \tanh^2(y)}} \right)} } \tanh(y).
\end{eqnarray}

\subsection{Case (b)}

Let us now consider {\bf{case (b)}}, with
\begin{eqnarray}
P_1^{\mu} &=& \left( E_1, |\mathbf{P}_1|\cos\phi,|\mathbf{P}_1|\sin\phi,0 \right)\\
P_2^{\mu} &=& \left( E_2, |\mathbf{P}_2|\sin\theta\cos\varphi,|\mathbf{P}_2|\sin\theta\sin\varphi,|\mathbf{P}_2| \cos\theta \right)
\end{eqnarray}
Therefore, we have (for on-shell gluons $|\mathbf{P}_1| = E_1$, $|\mathbf{P}_2| = E_2$)
\begin{eqnarray}
P_1\cdot P_2 = \frac{s}{2} = E_1 E_2 \left( 1 - \sin\theta\cos(\varphi-\phi) \right). 
\label{eq_b1}
\end{eqnarray}
The transverse momentum is now defined by
\begin{eqnarray}
\label{pT-b}
p_{T} = P_{1,z} + P_{2,z} = E_2\cos\theta.
\label{eq_b2}
\end{eqnarray}
Finally, the longitudinal momentum is given by
\begin{eqnarray}
p_{\parallel} &=& \sqrt{\left(   E_1\cos\phi + E_2\sin\theta\cos\varphi\right)^2 + \left( E_1\sin\phi + E_2\sin\theta\sin\varphi \right)^2}\nonumber\\
&=& \sqrt{E_1^2 + E_2^2\sin^2\theta + 2 E_1 E_2 \sin\theta \cos(\varphi-\phi)}.
\label{eq_b3}
\end{eqnarray}
The later defines the rapidity, given as previously discussed in case (a) by the relation
\begin{eqnarray}
\label{y-b}
\tanh y = \frac{p_{\parallel}}{P_0} = \frac{p_{\parallel}}{E_1 + E_2}.
\label{eq_b4}
\end{eqnarray}
Let us now define $\mathcal{C} = \cos(\varphi-\phi)$ as a free parameter. 


Summarizing, we end up with the system of equations for Case (b),
\begin{eqnarray}
E_1 E_2\left(1- \mathcal{C}\sin\theta \right) = \frac{s}{2},
\label{eq_Cb1}
\end{eqnarray}
\begin{eqnarray}
E_2\cos\theta = p_T,
\label{eq_Cb2}
\end{eqnarray}
\begin{eqnarray}
\tanh(y) = \frac{\sqrt{E_1^2 + E_2^2\sin^2\theta + 2 E_1 E_2 \mathcal{C}\sin\theta}}{E_1 + E_2}.
\label{eq_Cb3}
\end{eqnarray}
We will show that from this system, we can solve the set of kinematic variables $\left(E_1,E_2,\theta\right)$ in terms of $\left(s,p_T,y\right)$. For this purpose, we first notice that after Eq.~\eqref{eq_Cb1}, we have
\begin{equation}
2 E_1 E_2 \mathcal{C}\sin\theta = 2 E_1 E_2 - s,
\end{equation}
so inserting this into Eq.~\eqref{eq_Cb3}, we obtain
\begin{eqnarray}
\tanh(y) = \frac{\sqrt{E_1^2 + E_2^2 - E_2^2\cos^2\theta + 2 E_1 E_2 - s}}{E_1 + E_2} = \frac{\sqrt{(E_1 + E_2)^2 - s - p_{T}^2}}{E_1 + E_2}.
\end{eqnarray}
After elementary algebraic manipulations, the equation above can be solved to yield
\begin{eqnarray}
E_1 + E_2 = \sqrt{\frac{p_{T}^2 + s}{1 - \tanh^2(y)}}.
\label{eq_CbE1E2}
\end{eqnarray}
In what follows, we shall assume that $\mathcal{C}^2 = 1$ (this is equivalent to the cases $\varphi - \phi =0$ or $\varphi-\phi = \pi$ discussed in the main text). Then, we shall combine Eq.~\eqref{eq_Cb1} and Eq.~\eqref{eq_Cb2}, taking the squares and then adding them as follows
\begin{eqnarray}
E_2^2\mathcal{C}^2\sin^2\theta &=& \left( E_2 - \frac{s}{2 E_1} \right)^2 \\
E_2^2\cos^2\theta &=& p_{T}^2,
\end{eqnarray}
so that (for $\mathcal{C}^2 = 1$) we obtain
\begin{eqnarray}
\left( \cos^2\theta + \mathcal{C}^2\sin^2\theta \right)E_2^2 = E_2^2 = p_{T}^2 + \left( E_2 - \frac{s}{2 E_1} \right)^2.
\end{eqnarray}
After some elementary algebra, we get
\begin{eqnarray}
E_2 = \frac{s}{4 E_1} + E_1 \frac{p_{T}^2}{s}.
\end{eqnarray}
Substituting $E_2$ from the later equation into Eq.~\eqref{eq_CbE1E2}, we obtain
\begin{eqnarray}
E_1 + \frac{s}{4 E_1} + \frac{p_{T}^2}{s}E_1 = \sqrt{\frac{p_{T}^2 + s}{1 - \tanh^2(y)}} \equiv A
\end{eqnarray}
The later can be arranged as a quadratic equation for $E_1$, as follows
\begin{eqnarray}
\left( 1 + \frac{p_{T}^2}{s} \right)E_1^2 - A E_1 + \frac{s}{4} = 0,
\end{eqnarray}
whose solution is
\begin{eqnarray}
E_1 = \frac{A \pm \sqrt{A^2 - s - p_{T}^2}}{2\left( 1 + \frac{p_{T}^2}{s}  \right)}
= \sqrt{\frac{p_{T}^2 + s}{1 - \tanh^2(y)}} \frac{\left[1 \pm \tanh(y)\right]}{2\left( 1 + \frac{p_{T}^2}{s} \right)}
\end{eqnarray}
and canceling common factors, this reduces to
\begin{eqnarray}
E_1 = \sqrt{\frac{1 \pm \tanh(y)}{1 \mp \tanh(y)}}\frac{s}{2\sqrt{s + p_{T}^2}}
\end{eqnarray}
Finally, we solve for $E_2$, after Eq.~\eqref{eq_CbE1E2}
\begin{eqnarray}
E_2 &=& \sqrt{\frac{p_{T}^2 + s}{1 - \tanh^2(y)}} - E_1\nonumber\\
&=& \sqrt{\frac{p_{T}^2 + s}{1 - \tanh^2(y)}} - \sqrt{\frac{1 \pm \tanh(y)}{1 \mp \tanh(y)}}\frac{s}{2\sqrt{s + p_{T}^2}}\nonumber\\
&=& \frac{2 p_{T}^2 + s\left( 1 \mp \tanh(y) \right)}{2\sqrt{p_{T}^2 + s}\sqrt{1 - \tanh^2(y)} }.
\end{eqnarray}
Finally, the angle $\theta$ is obtained from the equation
\begin{eqnarray}
\mathcal{C}\tan\theta = \frac{2 E_1 E_2 - s}{2 p_{T} E_1}.
\end{eqnarray}

\section{The differential cross-section and rate}
\label{App:Cross_section}

Here we present the general expression for the differential cross-section involving the scattering of two incoming, massless particles. 

For the case of incoming gluons with arbitrary directions, we use the vector relations defining the group velocities, taking into account the relations for massless gluons $E_1 = |\mathbf{P}_1|$ and $E_2 = |\mathbf{P}_2|$,  
\begin{eqnarray}
\mathbf{v}_{1} &=& \frac{\partial E_1}{\partial\mathbf{P}_1} = \frac{\mathbf{p}_1}{|\mathbf{P}_1|} = \frac{\mathbf{P}_1}{E_1}\nonumber\\
\mathbf{v}_{2} &=& \frac{\partial E_2}{\partial\mathbf{P}_2} = \frac{\mathbf{P}_2}{|\mathbf{P}_2|} = \frac{\mathbf{P}_2}{E_2}
\end{eqnarray}
Therefore, in this general case,the denominator involved in the definition of the scattering differential cross-section can be expressed by
\begin{eqnarray}
E_1 E_2\left|\mathbf{v}_1 - \mathbf{v}_2  \right| = \left| E_2 \mathbf{P}_1 - E_1 \mathbf{P}_2 \right|.
\end{eqnarray}
Therefore, the differential cross section is then (for a single final particle, i.e. $f=\pi_0$, $f=\eta$, or $f=\eta'$, respectively)
\begin{eqnarray}
\label{diff-rate}
d\sigma (gg\rightarrow f) = \frac{1}{2 \left| E_2 \mathbf{P}_1 - E_1 \mathbf{P}_2 \right| } \left(  \frac{d^3 P_f}{(2\pi)^3} \frac{1}{2 E_f}\right) \left|  \mathcal{M}(P_1,P_2\rightarrow P_f)\right|^2 \left(  2\pi\right)^4 \delta^{(4)}\left( P_1 + P_2 - P_f \right)
\end{eqnarray}

From the definition of the differential cross-section above, the differential rate $d\tilde{R}(gg\rightarrow f)$ is obtained by multiplying by the net incident flux of scattering particles $J$ (particles per unit area, per unit time),
\begin{eqnarray}
d\tilde{R}(gg\rightarrow f) = J \frac{d\sigma}{d\Omega}d\Omega.
\end{eqnarray}
Since, concerning the microscopic scattering mechanism, the incoming flux is an (arbitrary) parameter, one can normalize the rate on the basis of $J = 1$~(particle per unit area, per unit time), such that in these units
\begin{eqnarray}
dR(gg\rightarrow f) \equiv \frac{1}{J}d\tilde{R}(gg\rightarrow f) = d\sigma(gg\rightarrow f),
\label{eq:diffrate}
\end{eqnarray}
the normalized rate and the differential cross-section possess the same numerical value.

\section{Integrals}
\label{App:Integrals}

In this Appendix, we summarize the analytical results for the phase-space integrals involved in
the calculation of the scattering amplitudes presented in the main text.

\subsection{Vacuum case}

\begin{eqnarray}
\mathcal{K}_0^{(0)} &=& \frac{1}{2 P_1\cdot P_2} \Bigg\{  \ln(z_1)\ln\left( \frac{1-z_1}{-z_1} \right) + \ln(z_2)\ln\left( \frac{1-z_2}{-z_2} \right)- Li_2\left( -\frac{(1-z_1)}{z_1} \right) - Li_2\left( -\frac{(1-z_2)}{z_2} \right) - \frac{\pi^2}{3}\Bigg\}
\end{eqnarray}
where
\begin{eqnarray}
z_{1,2} = \frac{1}{2} \pm \frac{1}{2}\sqrt{1 + \frac{2 M^2}{(P_1\cdot P_2)}}
\label{eq_z1z2}
\end{eqnarray}

\subsection{Weak magnetic field}

Here, it is convenient to remark the elementary identity, that follows from Eq.~\eqref{eq_z1z2}
\begin{equation}
z_1 z_2  = - \frac{M^2}{2 P_1 \cdot P_2}
\label{eq_z1z2prod}
\end{equation}

\begin{eqnarray}
\mathcal{K}_0^{(0)} &=& \frac{1}{2 P_1\cdot P_2} \left\{  \ln(z_1)\ln\left( \frac{1-z_1}{-z_1} \right) + \ln(z_2)\ln\left( \frac{1-z_2}{-z_2} \right)    \right.
\nonumber
\\ 
&& \left. - Li_2\left( -\frac{(1-z_1)}{z_1} \right) 
 - Li_2\left( -\frac{(1-z_2)}{z_2} \right) 
+ \frac{\pi^2}{3}\right\}
\end{eqnarray}

\begin{eqnarray}
\mathcal{J}_0^{(0)} = \frac{1}{4(P_1\cdot P_2)^2 (z_1 - z_2)}\left[ \frac{1}{z_1}\ln\left(\frac{1-z_1}{-z_1}\right) - \frac{1}{z_2}\ln\left(\frac{1-z_2}{-z_2}\right)\right]
\end{eqnarray}

\begin{eqnarray}
\mathcal{J}_0^{(1)} &=&  \frac{1}{2(P_1\cdot P_2)}\left[ \frac{1}{2(P_1\cdot P_2)(z_1 - z_2)}\ln\left( \frac{(1 - z_1)z_2}{(1 - z_2)z_1} \right) - \frac{1}{M^2} \right]
\end{eqnarray}
\begin{eqnarray}
\mathcal{J}_0^{(2)} &=&  -\frac{1}{2(P_1\cdot P_2)}\left[ \frac{-1}{2(P_1\cdot P_2)(z_1 - z_2)}\left[ z_1 \ln\left( \frac{1 -z_1}{-z_1} \right) - z_2 \ln\left( \frac{1 -z_2}{-z_2} \right)  \right] - \frac{1}{2 M^2}  \right]
\end{eqnarray}
\begin{eqnarray}
\mathcal{J}_0^{(3)} &=&  -\frac{1}{2(P_1\cdot P_2)}\left[ \frac{-1}{2(P_1\cdot P_2)(z_1 - z_2)}\left[ (z_1 - z_2) + z_1^2 \ln\left( \frac{1 -z_1}{-z_1} \right) - z_2^2 \ln\left( \frac{1 -z_2}{-z_2} \right)  \right] - \frac{1}{3 M^2}  \right]
\end{eqnarray}

\begin{eqnarray}
\mathcal{J}_1^{(0)} &=& 
 \frac{1}{\left(2 P_1\cdot P_2  \right) M^2} + \frac{1}{\left(  2 P_1\cdot P_2\right)^2(z_1 - z_2)}\ln\left[ \frac{(1 - z_1) z_2}{(1 - z_2)z_1} \right]
\end{eqnarray}

\begin{eqnarray}
\mathcal{J}_1^{(1)} &=&\frac{1}{\left(2 P_1\cdot P_2  \right)^2} \left\{ -\ln\left( \frac{1  -z_1}{-z_1} \right)\ln(z_1) -\ln\left( \frac{1  -z_2}{-z_2} \right)\ln(z_2) + Li_2\left(1 - \frac{1}{z_1}  \right) + Li_2\left(1 - \frac{1}{z_2}  \right) - \frac{\pi^2}{3}\right.\nonumber\\
&&\left.+ \frac{1}{z_1 - z_2}\left[ (1-z_1)\ln\left( \frac{1  -z_1}{-z_1}\right) - (1-z_2)\ln\left( \frac{1  -z_2}{-z_2}\right) \right]
\right\}
\end{eqnarray}
\begin{eqnarray}
\mathcal{J}_1^{(2)} &=&
\frac{1}{\left(2 P_1\cdot P_2  \right)^2} \left\{ 
 -z_1\ln\left(\frac{1-z_1}{-z_1}\right) -z_2\ln\left(\frac{1-z_2}{-z_2}\right) - 3 + \frac{M^2}{(-2 P_1\cdot P_2)(z_1 - z_2)}\ln\left[ \frac{(1-z_1)z_2}{(1-z_2)z_1}  \right]
\right\}
\end{eqnarray}

\begin{eqnarray}
\mathcal{J}_2^{(0)} &=& \frac{1}{\left( 2 P_1\cdot P_2  \right)^2}\left\{
\frac{ z_2\ln\left( \frac{1 - z_1}{-z_1} \right)  - z_1\ln\left( \frac{1 - z_2}{-z_2} \right)}{z_1 - z_2} - 1
\right\}
\end{eqnarray}
\begin{eqnarray}
\mathcal{J}_2^{(1)} &=& \frac{1}{\left( 2 P_1\cdot P_2  \right)^2} \left\{
1 + \frac{2}{(-2 P_1 \cdot P_2)(z_1 - z_2)}\left[  
\left( M^2 + (z_1-1)^2 P_1\cdot P_2 \right)\ln\left( \frac{1 - z_1}{-z_1}\right)\right.\right.\nonumber\\
&&\left.\left.
- 
\left( M^2 + (z_2-1)^2 P_1\cdot P_2 \right)\ln\left( \frac{1 - z_2}{-z_2}\right)
\right]
\right\}
\end{eqnarray}

\end{document}